\documentclass[letterpaper,twocolumn,10pt]{article}
\usepackage{static/style/usenix-2020-09}

\usepackage{booktabs} 
\usepackage{multirow} 
\usepackage{graphicx} 
\usepackage{enumitem}

\usepackage{tikz}
\usepackage{amsmath}
\usepackage{url}

\usepackage{filecontents}

\usepackage{subcaption}
\usepackage{authblk}
\usepackage{breakurl}

\makeatletter
\renewcommand{\@makefntext}[1]{%
  \setlength{\parindent}{0pt}%
  \setlength{\leftskip}{0em}%
  \noindent\hbox{\@thefnmark~}#1}
\makeatother

\begin{document}

\date{}

\title{PipeBoost: Resilient Pipelined Architecture for Fast Serverless LLM Scaling}

\author{
Chongpeng Liu \hspace{0.5em} 
Xiaojian Liao$^{*}$ \hspace{0.5em} 
Hancheng Liu \hspace{0.5em}
Limin Xiao \hspace{0.5em}
Jianxin Li$^{*}$ \hspace{0.5em}
}
\affil{Beihang University}
\maketitle

\renewcommand{\thefootnote}{}
\footnotetext{$^{*}$Corresponding author: \{liaoxj, lijx\}@buaa.edu.cn}

\begin{abstract}
This paper presents PipeBoost, a low-latency LLM serving system for multi-GPU (serverless) clusters, which can rapidly launch inference services in response to bursty requests without preemptively over-provisioning GPUs. 
Many LLM inference tasks rely on the same base model (e.g., LoRA). To leverage this, PipeBoost introduces fault-tolerant pipeline parallelism across both model loading and inference stages. This approach maximizes aggregate PCIe bandwidth and parallel computation across GPUs, enabling faster generation of the first token.
PipeBoost also introduces recovery techniques that enable uninterrupted inference services by utilizing the shared advantages of multiple GPUs. 
Experimental results show that, compared to state-of-the-art low-latency LLM serving systems, PipeBoost reduces inference latency by 31\% to 49.8\%. 
For certain models (e.g., OPT-1.3B), PipeBoost achieves cold-start latencies in the range of a few hundred microseconds.
\end{abstract}

\section{Introduction}
Large Language Models (LLMs) have recently become popular and crucial for applications such as chatbots (e.g., ChatGPT) \cite{IntroducingChatGPT}, search engines \cite{IntroducingNewBing}, code assistants (e.g., Copilot) \cite{GitHubCopilotYoura}, and AI-driven decision making \cite{yaoTreeThoughtsDeliberate2023,wang2024boosting}, where low latency and high reliability are critical for serving performance.
LLMs require significant GPU memory space due to their massive number of parameters, posing a substantial challenge to large-scale LLM serving and contributing to its high cost \cite{InformationBeautiful}. 
For example, an OPT-66B model~\cite{zhangOPTOpenPretrained2022} requires 132GB A100 GPUs.

Deploying LLMs on serverless GPU clusters holds immense potential, enabling service providers to efficiently multiplex LLMs across GPUs \cite{huangENOVAAutoscalingCosteffective2024}. 
Users submit inference requests and pay per use, without needing to manage the complexities of LLM deployment. 
This approach not only increases GPU resource utilization but also significantly lowers the barrier for users to access and leverage LLM capabilities.

However, serverless workloads are often unpredictable and face frequent load spikes \cite{mohanAgileColdStarts2019,sahraeiXFaaSHyperscaleLow2023}, which introduce challenges to serving LLMs at scale.
Provisioning sufficient GPUs for potential load spikes is expensive, especially given the substantial memory demands of LLMs; GPUs reserved for peak traffic periods remain underutilized during off-peak times \cite{gaoEmpiricalStudyLow2024}.
Starting LLM services on-demand retains the auto-scaling benefits of serverless computing but faces high latency caused by cold starts (i.e., loading models from CPU memory or SSDs and initiating inference from scratch).
Our study (\S\ref{sec:motivation}) shows that even state-of-the-art systems like ServerlessLLM \cite{fuServerlessLLMLowLatencyServerless2024}, which store model checkpoints in local CPU memory, require around 90\% of the time for model loading during a cold start, delaying the return of the first token to users.

The conventional wisdom is that a GPU can not start LLM inference until it fully loads model parameters.
In this paper, we demonstrate that this is not the case for serverless LLM due to the base model sharing property in LLM serving, which can be prevalent in practice and is overlooked by existing systems.
For instance, LLM service providers like OpenAI often deploy their proprietary LLMs (e.g., GPT-4~\cite{openaiGPT4TechnicalReport2024}) at scale to power AI-driven applications such as ChatGPT. 
Additionally, training an LLM from scratch is prohibitively expensive and requires extensive engineering expertise~\cite{rajbhandariZeROMemoryOptimizations2020}, making it inaccessible to many developers \cite{deepseek-aiDeepSeekV2StrongEconomical2024,brownLanguageModelsAre2020,jiangMegaScaleScalingLarge}. 
Instead, these developers commonly fine-tune existing base models for specific domains.
LoRA (Low-Rank Adaptation) \cite{huLoRALowRankAdaptation2021} is a widely adopted parameter-efficient fine-tuning technique. 
It keeps the base model intact and introduces smaller matrices to encode the delta adjustments, enabling efficient customization without modifying the original base model.

We introduce PipeBoost, a serverless LLM serving system that achieves low-latency cold starts (\S\ref{sec:design}).
Leveraging the base mode sharing property, PipeBoost breaks the sequential execution order of LLM loading and inference stages, allowing them to be performed in a pipelined fashion.
The key idea lies in grouping LLMs with the same base model onto the same GPU server and introducing fault-tolerant pipeline parallelism across both the loading and inference stages.
Specifically, it includes the following three key designs:

\noindent
\textbf{(1) Pipeline-Parallel Model Loading} (\S\ref{sec:pp_model_loading}). 
PipeBoost reorders the model layers loaded by each GPU during the model loading stage, allowing GPUs to load distinct parts of the same base model and LoRA adapters. 
This design eliminates redundant model layer transfers along the critical path, efficiently utilizing aggregated PCIe and memory bandwidth, and enabling the system to reach a ready-to-infer state quickly.

\noindent
\textbf{(2) Pipeline-Parallel Model Inference} (\S\ref{sec:pp_model_inference}).
When model layers across GPUs form a complete model, PipeBoost initiates the inference, leveraging pipeline parallelism, merged LoRA, and epoch-based adapter switching techniques to handle bursty requests while concurrently loading the remaining layers. 
This design exploits pipeline parallelism to rapidly activate multi-GPU parallel processing capabilities, effectively managing sudden traffic surges.
PipeBoost also features a seamless strategy-switching mechanism, allowing inference processes to transition to alternative parallel strategies.
This flexibility is crucial as different GPU servers may be optimized for specific use cases (e.g., long-sequence inference).
Moreover, pipeline parallelism, while advantageous in launching LLM services, can perform worse compared to single-GPU inference (where each GPU independently performs inference) under high request arrival rates due to the significant communication overhead between GPUs.

\noindent
\textbf{(3) Pipeline-Parallel Recovery} (\S\ref{sec:design_fault}).
Pipeline parallelism is more fragile to failures. 
Crash in a single LLM instance can impact all instances and requests within the pipeline.
PipeBoost introduces model layer reassignment and KV cache reconstruction techniques to swiftly restore the pipeline-parallel model loading and inference process.
This design mitigates the drawbacks of pipeline parallelism, improving system reliability, reducing the recovery time and ensuring predictable cold-start latencies.

We implement PipeBoost based on PyTorch~\cite{paszkePyTorchImperativeStyle2019} and Transformers~\cite{wolf-etal-2020-transformers}, supporting a wide range of LLMs, including OPT, Falcon, and Mistral.
Our experimental evaluation (\S\ref{sec:figs}) compares PipeBoost against existing systems such as Transformers and ServerlessLLM.
The results demonstrate that PipeBoost reduces the time to first token by 57\%–84\% and 30\%–47\% compared to Transformers and ServerlessLLM, respectively.
Notably, PipeBoost can initialize OPT-1.3B and generate the first token within 1 second, with the model loading time costing only 0.64 seconds.
Recovery experiments further show that PipeBoost reduces recovery time by 51\% compared to traditional methods that restart LLM instances.

In summary, we make the following contributions.
\begin{itemize}[noitemsep,topsep=0pt,parsep=0pt,partopsep=0pt]
    \item We analyze the limitations of existing serverless LLM serving systems and highlight the overlooked characteristic of base model sharing.
    \item We propose PipeBoost, introducing the Fault-Tolerant Pipeline Parallelism to maximize the parallel computing and data transfer capabilities of multi-GPU servers.
    \item We conduct extensive experiments to demonstrate that PipeBoost achieves faster startup and recovery latencies compared to state-of-the-art systems.
\end{itemize}

\section{Background}
This section first introduces the background of LLM serving, followed by an explanation of the cold-start challenges when deploying LLMs in serverless environments.

\subsection{LLM Serving}\label{sec:background_llm}
LLM serving primarily consists of two parts: model loading and model inference.
In the model loading stage, in addition to initializing the runtime (e.g., Python) and other prerequisites, it involves two steps.
First, a model checkpoint is loaded from a persistent storage device (such as local NVMe SSDs or remote storage like S3) into CPU memory and deserialized into an executable format. 
Next, if inference is performed on the GPU, the model checkpoint or parameters must be transferred from CPU memory to GPU memory.
Once the model is loaded, the LLM serving can perform inference.

LLM inference consists of two stages: prefill and decode. 
Taking a transformer-based model as an example, the model comprises multiple layers, where the output of one layer serves as the input to the next.

During the prefill stage, the inference engine inputs initial tokens into the first layer of the model. 
Each layer performs two key operations. 
First, it computes the relationships among input tokens using a self-attention mechanism. 
Then, it applies a feed-forward network (FFN) to produce the output for the current layer. 
This output is passed to the subsequent layer, where similar computations are performed. 
The final layer of the model generates the first token to be returned to the user, ending the prefill stage and starting the decode stage.

In the decode stage, the inference engine generates tokens iteratively, processing only the most recent token at each step until reaching either an end-of-sequence token or the maximum sequence length. KV cache technology eliminates the need to reprocess the full input prompt or prior tokens by storing intermediate key (K) and value (V) matrices in GPU memory. Each model layer maintains its own KV cache, enabling efficient self-attention computations and reducing latency.

Low-Rank Adaptation (LoRA)~\cite{huLoRALowRankAdaptation2021} introduces low-rank trainable matrices, referred to as adapters, into the model's architecture, specifically in the weight matrices of layers like attention and FFNs in transformer models.
It reduces the number of parameters that need to be updated during fine-tuning while still enabling the model to adapt effectively to new tasks.
In LoRA, the original weight matrix \textit{W} remains fixed during training, and only the low-rank matrices \textit{A} and \textit{B} (with rank \textit{r} and dimensions \textit{d} $\times$ \textit{r} and \textit{r} $\times$ \textit{d}, respectively) are updated. 
Since the rank \textit{r} is substantially smaller than the dimension of \textit{W} (i.e., \textit{d}), the memory footprint of the LoRA adapter is significantly reduced compared to the original model.

\subsection{Serverless LLM}
Serverless computing has gained popularity in cloud computing due to its exceptional elasticity. 
Many companies, including Amazon, Azure, and Hugging Face, have implemented serverless LLM services in their cloud platforms \cite{AmazonSageMaker,AzureFunctionsServerlessa,HuggingFaceGenerative}. 
Users only need to submit functions related to LLM inference, without the need to rent containers or virtual machines and deploy their own inference systems as in traditional methods, greatly reducing the development burden on users.

Serverless computing requires efficient scaling capabilities. 
In cloud workloads (e.g., Azure Functions \cite{AzureFunctionsServerlessa}), load spikes are common and often unpredictable. 
To handle these spikes, mainstream serverless platforms typically use two approaches: over-provisioning and on-demand cold starts.

In the over-provisioning approach, the serverless platform reserves GPUs and pre-launches related services in advance. 
However, this method leads to significant GPU resource wastage. 
First, because the peak of burst traffic is generally unpredictable, it is difficult to determine the necessary number of reserved GPUs. 
Second, the reserved GPUs result in wasted GPU compute power and memory, and this waste is exacerbated in LLM inference scenarios, as LLM parameters can occupy up to 800 GB of memory (e.g., Llama 3.1 405B~\cite{IntroducingLlama31}), often requiring multiple GPUs for complete loading.

In the on-demand cold start approach, when existing GPUs cannot handle the current traffic load (i.e., when a load spike occurs), the serverless scheduler allocates additional GPUs to the LLM inference tasks. 
At this point, the LLM inference service must go through the model loading stage before it can begin inference, a process known as LLM cold start.

LLM cold starts can be prevalent in real-world settings.
As demonstrated in the ServerlessLLM \cite{fuServerlessLLMLowLatencyServerless2024}, if the distribution of LLM inference load follows the Azure trace \cite{shahradServerlessWildCharacterizing2020}, over 40\% of functions experience a cold start rate exceeding 25\%, and approximately 25\% of functions face a cold start rate greater than 60\%, within a 5-minute keep-alive interval.
\begin{figure}[t] 
  \centering 
  \includegraphics[width=\linewidth]{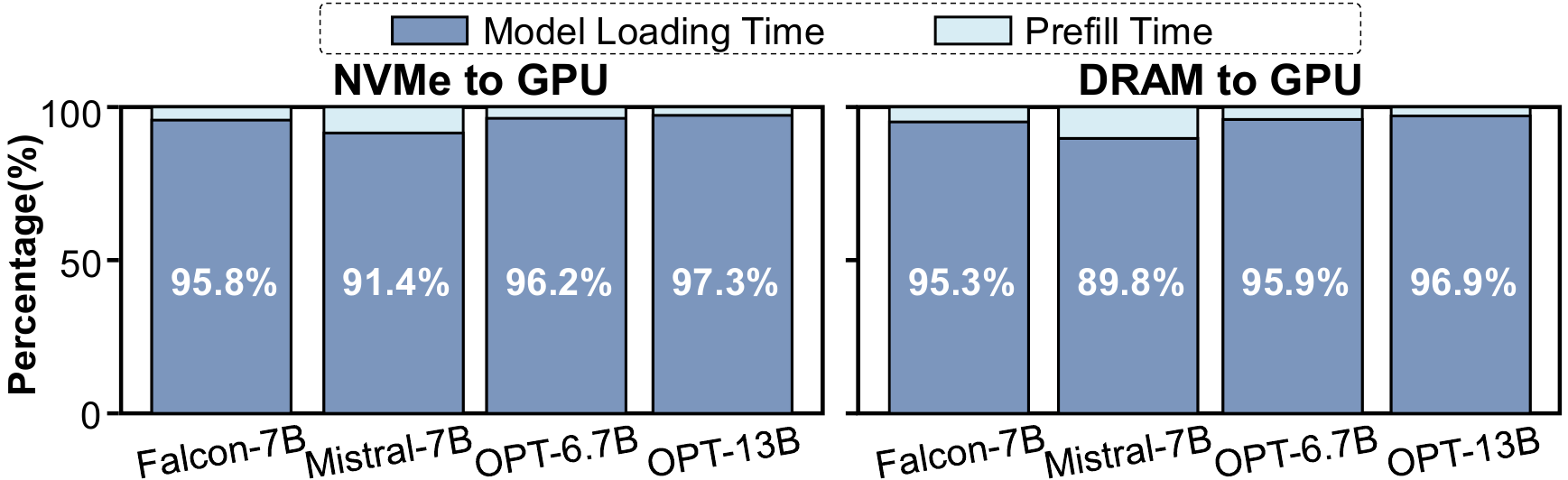} 
  \caption{\textbf{Overhead of LLM model loading.}}
  \label{fig:moti_loading_overhead} 
\end{figure}

\begin{figure*}[t] 
  \centering 
  \includegraphics[width=\linewidth]{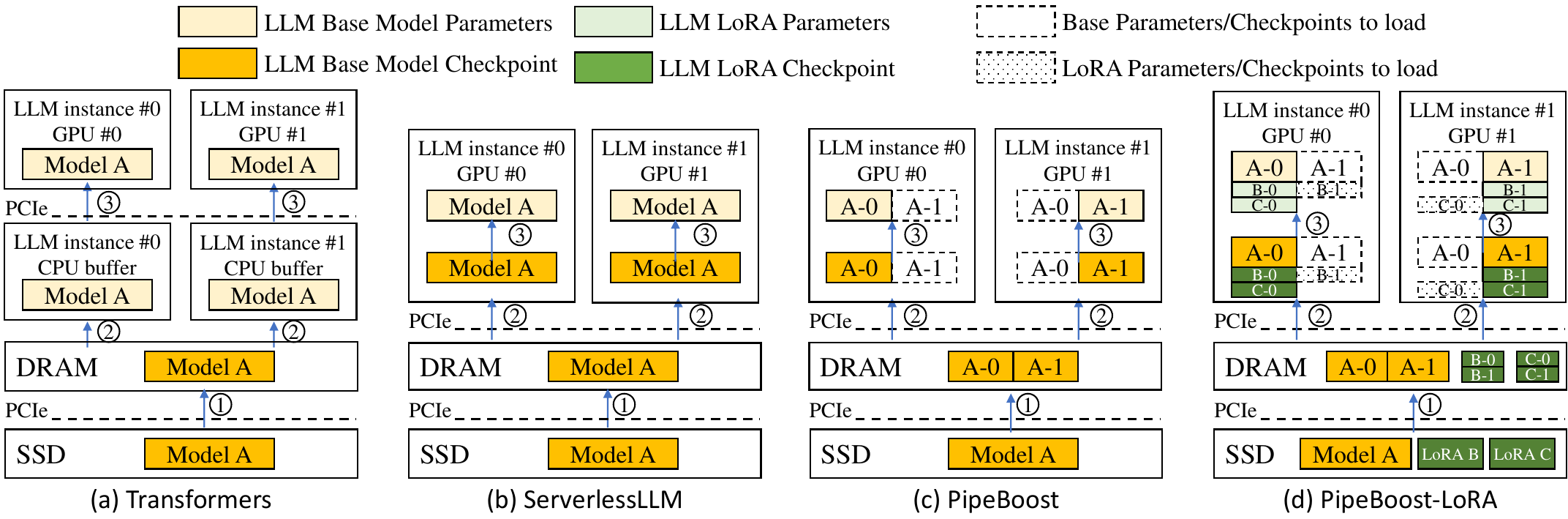} 
  \caption{\textbf{Different methods of LLM model loading on multiple GPUs.} Transformers and ServerlessLLM require the entire set of LLM parameters to be fully loaded into GPU memory before inference can begin. In contrast, PipeBoost allows different GPUs to load distinct portions of the LLM layers, enabling inference to start without waiting for the full model to be loaded.}
  \label{fig:ftpp_loading} 
\end{figure*}

\section{Motivation}\label{sec:motivation}
The severe performance overhead of serverless cold starts has been a known challenge in big data tasks \cite{fuServerlessLLMLowLatencyServerless2024}. 
This overhead is even more pronounced in LLM inference tasks, as LLM serving systems require loading a large number of parameters. 
If these parameters are stored on a slower remote storage system (e.g., S3), loading delays can exceed 60 seconds, which is intolerable in interactive services requiring low latency.

To mitigate the impact of LLM cold starts, ServerlessLLM \cite{fuServerlessLLMLowLatencyServerless2024} stores model parameters in local CPU memory and high-performance NVMe SSDs, leveraging the high-speed PCIe bus to significantly reduce LLM loading time.
Although powerful, the model loading time in ServerlessLLM is significantly higher than the latency of the prefill stage, which could severely impact user experience when handling burst traffic.

\subsection{Overhead of LLM Cold Starts}

This section quantifies the LLM cold start overhead in ServerlessLLM for generating the first token. 
Similar to mainstream LLM inference frameworks such as Transformer \cite{wolf-etal-2020-transformers} and vLLM \cite{kwonEfficientMemoryManagement2023c}, ServerlessLLM requires deserialization and loading parameters into GPU memory before inference can begin (i.e., entering the prefill stage). 
Experiments were conducted on a 2-GPU 40GB A100 machine with 512GB of DDR4 memory and a 2TB NVMe SSD. 
Detailed experimental environments are provided in \S\ref{sec:eval_starup}. 
The experiment tests mainstream open-source LLMs like Falcon, Mistral and OPT. 
The models were placed either in CPU memory or on NVMe SSD, and measurements were taken for the latency ratio of model loading to the prefill stage for generating the first token.

Figure~\ref{fig:moti_loading_overhead} reveals that even with local high-performance NVMe storage, the model loading time still accounts for an average of 95.2\% of the overall time. 
Even when the model resides in CPU memory, the time taken to load the model from the CPU to the GPU still constitutes 94.6\% of the total.
The experimental results demonstrate that once an LLM cold start occurs, the latency for generating the first token will increase by at least 9.8$\times$, severely impacting user experience.

\subsection{Observation and Analysis}\label{sec:motivation_observation}

Base model sharing across multiple GPU inference instances, a unique property in serverless LLM inference, is common in private cloud and LLM serving platforms.  
For example, OpenAI's ChatGPT service uses its own GPT-4o or GPT-4o mini model \cite{IntroducingChatGPT}, while Moonshot AI’s LLM inference service utilizes its proprietary Kimi model \cite{qinMooncakeKVCachecentricDisaggregated2024a}. 
Even in deployments with multiple models (such as a mix of LLaMa and OPT), inference clusters often contain numerous GPU instances running inference with the same model \cite{liAlpaServeStatisticalMultiplexing2023a}.

Another observation is that even when models are customized based on different professional domains or user preferences, these custom models typically share a common base model. 
Training a model from scratch is prohibitively expensive \cite{huLoRALowRankAdaptation2021}, so techniques like LoRA (details in \S\ref{sec:background_llm}) are often used to fine-tune an existing base model by adjusting parameters to suit specialized applications.
LoRA adapters are notably compact, taking up only a small fraction of base model parameters—for example, just 0.01\% in GPT-3 175B \cite{huLoRALowRankAdaptation2021}.

In summary, on an inference server equipped with multiple GPUs, it is common for multiple LLM instances to share the same full model or base model. 
Current serverless LLM inference systems do not account for this, which can lead to inefficient PCIe bandwidth utilization during the auto-scaling of LLM instances.
We illustrate this issue by taking state-of-the-art serverless LLM inference systems including Transformers and ServerlessLLM as examples (Figure~\ref{fig:ftpp_loading}). 
For simplicity, we assume a server with only two GPUs, with the process extending similarly for more GPUs.

Transformers first load model A from the SSD into CPU DRAM. 
Next, the model checkpoint is converted into an executable inference format (i.e., LLM parameters). 
It is important to note that an LLM instance is launched on each GPU, resulting in two copies of the LLM parameters residing on the CPU side. 
Subsequently, the LLM parameters are separately transferred to each GPU for inference execution.

The LLM startup process in ServerlessLLM is similar to that of Transformers, 
with the only difference being that the LLM checkpoint is first transferred to the GPU, where the GPU handles the conversion from checkpoint to parameters.

A typical GPU server often has high CPU memory bandwidth, with multiple memory channels providing aggregate bandwidth exceeding hundreds of GB/s. 
Generally, each GPU device is connected to the CPU through a dedicated PCIe slot (e.g., PCIe 4.0 x16) by up to 32~GB/s, allowing for high-speed transfer of model checkpoints or parameters between CPU memory and GPU HBM. 
However, the model loading methods of Transformer and ServerlessLLM do not efficiently leverage the high bandwidth of CPU memory and PCIe links.
In a given time window, all GPUs may be reading the same portion of the model checkpoint or parameters (e.g., all 8 GPUs reading the first 1~GB portion of the model), which results in the redundant transmission of large amounts of duplicate data along the critical I/O path that requires low latency.
Even worse, GPUs cannot begin inference until the entire model's parameters are fully loaded, and the GPU's computation capability is thus left underutilized.

\begin{figure}[t] 
  \centering 
  \includegraphics[width=\linewidth]{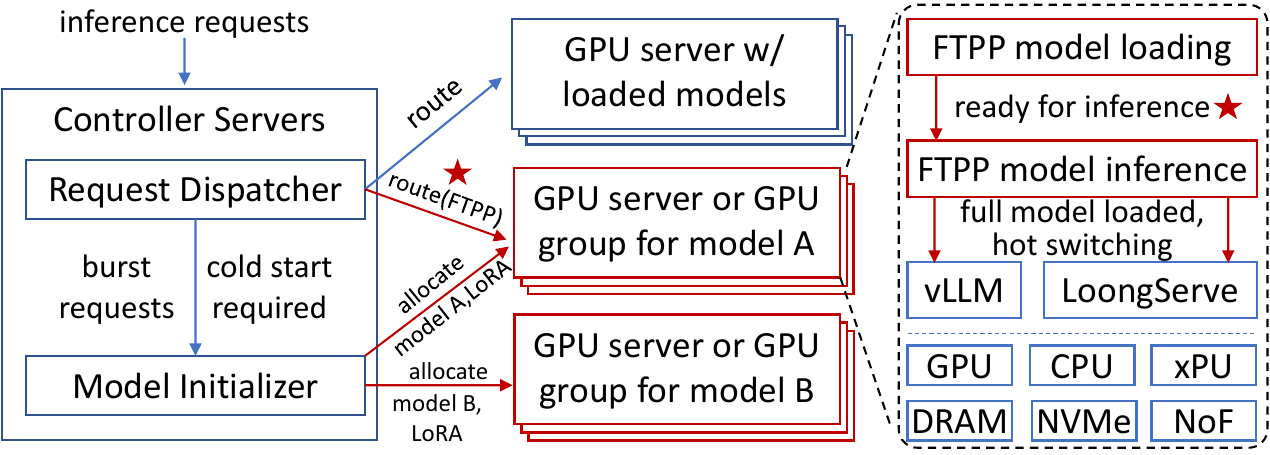} 
  \caption{\textbf{PipeBoost overview.} Red rectangles and arrows are new designs introduced by PipeBoost. Model checkpoints or parameters are stored on CPU DRAM, local NVMe SSDs or remote storage, e.g., NVMe over Fabrics (NoF), and are loaded into GPU during LLM inference cold starts.}
  \label{fig:pipeboost_overview} 
\end{figure}

\section{PipeBoost}\label{sec:design}
To address the high model loading latency discussed in \S\ref{sec:motivation}, we propose PipeBoost, a low-latency LLM serving system for serverless computing. 

\subsection{Overview}
Similar to traditional GPU serverless systems, PipeBoost consists of Controller Servers and GPU Servers (Figure~\ref{fig:pipeboost_overview}). 
The Controller Servers handle requests and forward them, via the Request Dispatcher, to servers where the LLM inference service is initialized.
In cases of sudden traffic spikes, the Model Initializer launches LLM inference instances on additional GPU servers to handle subsequent requests.

Inspired by the observation that multiple GPUs can share a base model in the cluster (\S\ref{sec:motivation_observation}), PipeBoost categorizes models and ensures that LLM inference instances using the same class of models are launched on the same GPU server. 
Notably, the model parameters do not need to be identical; models based on the same base model but using different LoRA adapters are also scheduled to launch on the same GPU server.

Within each GPU server, PipeBoost introduces the Fault-Tolerant Pipeline Parallelism (FTPP) technique to accelerate the startup process of inference services. Specifically:
\begin{itemize}[noitemsep,topsep=0pt,parsep=0pt,partopsep=0pt]
\item \textbf{Pipeline-Parallel Model Loading} (\S\ref{sec:pp_model_loading}): PipeBoost partitions the model checkpoints by layers (i.e., inter-operator partition) and loads it across GPUs, enabling the GPU server to reach a request-inference state as quickly as possible, achieving low-latency cold starts.
\item \textbf{Pipeline-Parallel Model Inference} (\S\ref{sec:pp_model_inference}): PipeBoost leverages the parallelism of multiple GPUs to handle burst requests while asynchronously loading the remaining layers of the model onto each GPU. This process continues until each GPU holds a complete copy of the model parameters. Once the full model is loaded, PipeBoost can either maintain the pipeline-parallel strategy or perform a seamless hot-switching to an existing inference strategy optimized for specific use cases.
\item \textbf{Fault Tolerance} (\S\ref{sec:design_fault}): PipeBoost introduces rapid recovery techniques to minimize the performance impact of partial LLM instance failures on pipeline inference. By dynamically adjusting the number of model layers loaded on each GPU and reconstructing KV caches on the fly, PipeBoost maximizes the advantages of pipeline parallelism in handling bursty requests.
\end{itemize}

While FTPP is primarily designed for GPU inference, it is also adaptable to CPU and other heterogeneous computing devices compatible with PyTorch. 
For clarity, this paper focuses on describing FTPP in the context of GPU inference. 

\subsection{Pipeline-Parallel Model Loading}\label{sec:pp_model_loading}
FTPP optimizes the transfer of model checkpoints by reordering the portions sent to different GPUs, effectively eliminating redundant data transfers from CPU memory to GPU HBM in the critical path. 

\subsubsection{Base Model Loading}
Inspired by the observation that multiple GPUs share a base model, FTPP prioritizes the transfer of non-duplicated model checkpoints during the inference service startup to maximize the effective utilization of memory and PCIe bandwidth. 
As shown in Figure~\ref{fig:ftpp_loading}(c), PipeBoost first reads the model checkpoint from SSD into DRAM or uses a model checkpoint already residing in DRAM. 
Next, PipeBoost splits the model checkpoint in DRAM into \textit{N} parts, with each part corresponding to a GPU, where \textit{N} represents the number of GPUs. 
PipeBoost then transfers each part of the checkpoint in parallel to the GPUs.
Crucially, during this step, PipeBoost utilizes the available PCIe bandwidth for transferring non-duplicated checkpoint parts, e.g., GPU 0 reads model A-0 while GPU 1 reads A-1. 
Finally, each GPU independently converts its checkpoint part into parameters. 
At this point, the inference service is successfully initialized and ready to serve.
While inference is being executed on GPUs (which will be detailed in \S\ref{sec:pp_model_inference}), the remaining portions of the model are progressively loaded into GPU HBM. For instance, model A-1 is loaded onto GPU 0, and model A-0 is loaded onto GPU 1.

\subsubsection{LoRA Adapter Loading}
PipeBoost also supports efficient loading of LoRA models. Suppose GPU 0 and GPU 1 need to load two models that share the same base model A but use different LoRA adapters B and C.
The loading process for the base model follows the previously described approach (as shown in Figure~\ref{fig:ftpp_loading}(c)). 
The loading of LoRA adapters is similar to that of the base model, where each adapter is partitioned, and the segmented parts are distributed across GPUs.
Notably, to handle requests requiring different LoRA adapters, each GPU also loads an additional portion of the other adapter. For example, GPU 0 loads C-0, while GPU 1 loads B-1. 
Since LoRA adapters are relatively lightweight, typically consuming only a fraction of the size of the base model (e.g., 0.01\%), the time and memory overhead for loading these adapters is negligible.
Once both the base model and LoRA adapters are ready, PipeBoost begins serving inference requests while asynchronously loading the remaining parts of the model into GPUs. 
After all model parameters have been fully loaded, the additional LoRA adapter segments can be discarded to free up GPU memory. 
For instance, GPU 0 discards C-0, and GPU 1 discards B-1.

\begin{figure*}[t] 
  \centering 
  \includegraphics[width=\linewidth]{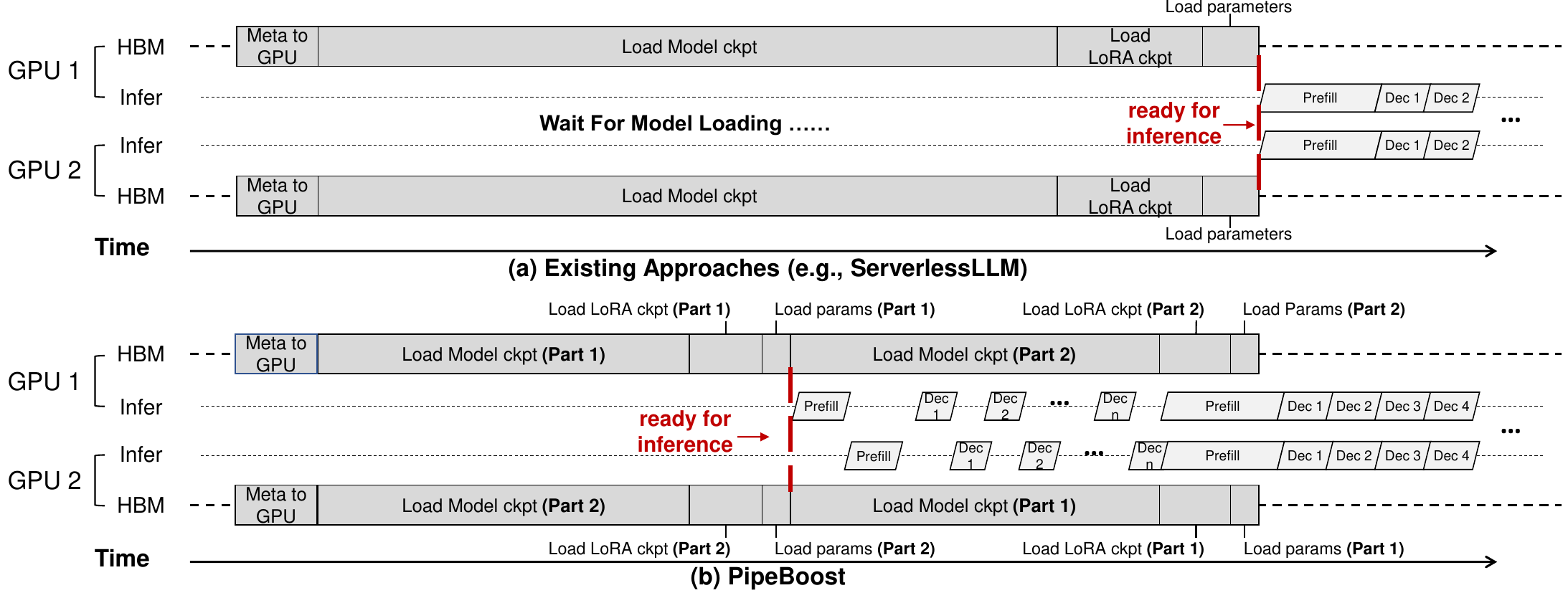} 
  \caption{\textbf{Different methods of LLM model inference during LLM cold starts on multiple GPUs.}}
  \label{fig:ftpp_inference} 
\end{figure*}

\subsection{Pipeline-Parallel Model Inference}\label{sec:pp_model_inference}
When the model layers already loaded across GPUs can be combined into a complete model, PipeBoost immediately utilizes pipeline parallelism to begin inference. 

\subsubsection{Base Model Inference}
As shown in Figure~\ref{fig:ftpp_inference}(b), all inference requests begin the prefill computation phase on GPU 0 which contains the first half portion of layers, while the remaining layers of the base model and LoRA adapters are concurrently loaded onto each GPU. 
Once GPU 0 completes computations, PipeBoost transfers the intermediate results to GPU 1 via NVLink or PCIe, where the remaining prefill computations are finalized. 
At the end of the prefill phase, the first token is generated and returned to both users and GPU 0.
The GPU 0 then begins the decode computation phase. Each decode step employs a similar pipeline parallelism approach as the prefill phase, continuing until a termination token is encountered.

\subsubsection{Merged LoRA Adapter Inference}\label{sec:design_pp_inference_lora}

PipeBoost employs the merged LoRA approach for pipeline inference, where the parameters of the LoRA adapter are merged back into the base model prior to inference, forming a full model. 
Inference is then conducted on the full model, following the same process outlined in the previous paragraph.
We chose the merged LoRA approach for the following reasons.
First, compared to unmerged LoRA, merged LoRA eliminates additional computational overhead (which accounts for approximately 38\% of the original computation \cite{wuDLoRADynamicallyOrchestrating2024}), enabling faster generation of the first token during request surges. 
Second, PipeBoost groups and batches requests with different LoRA adapters, significantly reducing the overhead caused by switching between LoRA adapters. 
Finally, PipeBoost is designed to eventually converge to a strategy where each GPU performs independent inference using a single LoRA adapter (reasons in \S\ref{sec:design_strategy_switch}). 
The merged LoRA approach allows for a smoother transition to this strategy.

\begin{figure}[t] 
  \centering 
  \includegraphics[width=\linewidth]{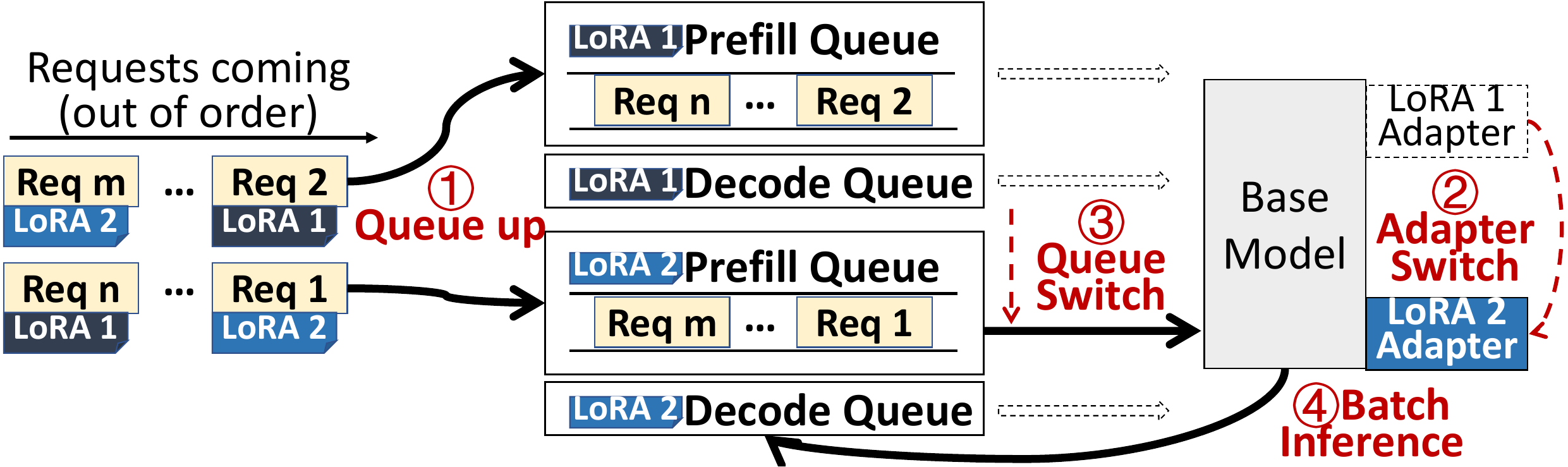} 
  \caption{\textbf{Details of epoch-based adapter switching.}}
  \label{fig:epoch-adapeter-switch} 
\end{figure}

In PipeBoost, requests associated with different LoRA adapters flow through all GPUs, potentially leading to frequent adapter switching. 
To address this issue, PipeBoost employs an epoch-based adapter switching technique (Figure~\ref{fig:epoch-adapeter-switch}).
First, PipeBoost classifies requests and groups those belonging to the same adapter into the same queue.
Second, leveraging the merged LoRA approach, PipeBoost prioritizes the scheduling of batches corresponding to the currently activated adapter. 
At regular intervals, PipeBoost switches adapters to serve requests from other batches. 
Notably, due to the use of pipeline parallelism, adapter switching across GPUs does not occur simultaneously but sequentially. 
Each GPU switches adapters only after completing the batch received from the preceding GPU in the pipeline.

\subsubsection{Seamless Strategy Switching}\label{sec:design_strategy_switch}

Once all GPUs in a server have fully loaded the complete model, PipeBoost can seamlessly switch to other inference strategies, such as vLLM's single-GPU independent inference. 
The switching process is straightforward: batches of requests submitted after the switching point are executed using the new inference parallelism strategy.

We propose transitioning PipeBoost from pipeline parallelism to other strategies for the following reasons.
First, while pipeline parallelism excels in quickly initializing inference services to handle bursty requests, its advantages diminish once all GPUs have fully loaded the complete model. 

\begin{figure}[t] 
  \centering 
  \includegraphics[width=0.9\linewidth]{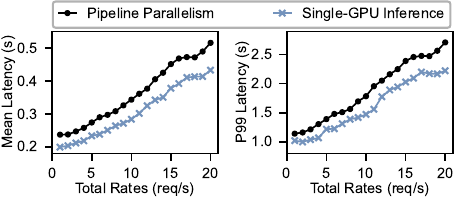} 
  \caption{\textbf{Necessity of Seamless Strategy Switching.} Single-GPU inference, where each GPU independently performs inference, outperforms pipeline-parallel model inference.}
  \label{fig:necessity_of_switch_to_single_gpu} 
\end{figure}

Figure~\ref{fig:necessity_of_switch_to_single_gpu} shows that , independent of the total request arrival rates, single-GPU inference, where requests are evenly distributed across GPUs and each GPU independently executes inference, consistently achieves significantly lower latency compared to pipeline-parallel model inference. 
Furthermore, the gap between the two widens as the total rates increase.
This phenomenon occurs because pipeline parallelism introduces communication overhead for transferring hidden states between GPUs. 
As the request density increases, this overhead scales accordingly, leading to a more pronounced latency difference.
Hence, PipeBoost opts to switch the inference strategy to single-GPU inference.

Second, different workloads have distinct characteristics, which may require tailored optimization strategies. 
For example, in handling long-sequence input, sequence parallelism offers superior performance compared to other inference parallelism strategies, as shown in \cite{wuLoongServeEfficientlyServing2024a}. 
In such scenarios, repurposing the GPU server for long-sequence tasks may be a more suitable approach.

\subsection{Fault Tolerance}\label{sec:design_fault}
This section first explains the necessity of providing efficient crash recovery support for pipeline-parallel model loading and inference, and then details the specific recovery process.

\subsubsection{Need For Fault Tolerance}\label{sec:need_for_fault_tolerance}
LLM instance crashes are common in large-scale GPU clusters, often caused by GPU hardware failures~\cite{mohanCheckFreqFrequentFineGrained2021}, driver errors, or bugs in the CPU-side initialization program~\cite{grattafioriLlama3Herd2024,huCharacterizationLargeLanguage2024a}. 

A typical approach to address such crashes is for the Controller Servers to select another available GPU and reload the model~\cite{thorpeBambooMakingPreemptible2023a,wangGEMINIFastFailure2023a,lianUniversalCheckpointingEfficient2024}. 
However, this method is inefficient and results in unpredictable model loading and inference latency.

Another consideration is isolation. 
In existing methods like Transformers and ServerlessLLM, each GPU independently loads and serves the model. 
A single GPU failure does not impact requests to other GPUs. 
In contrast, PipeBoost requires GPUs to collaborate during the model loading and the initial stage of inference. 
If a single GPU fails and the issue is not handled, it could affect all requests directed to the GPU server.

\begin{figure}[t] 
  \centering 
  \includegraphics[width=\linewidth]{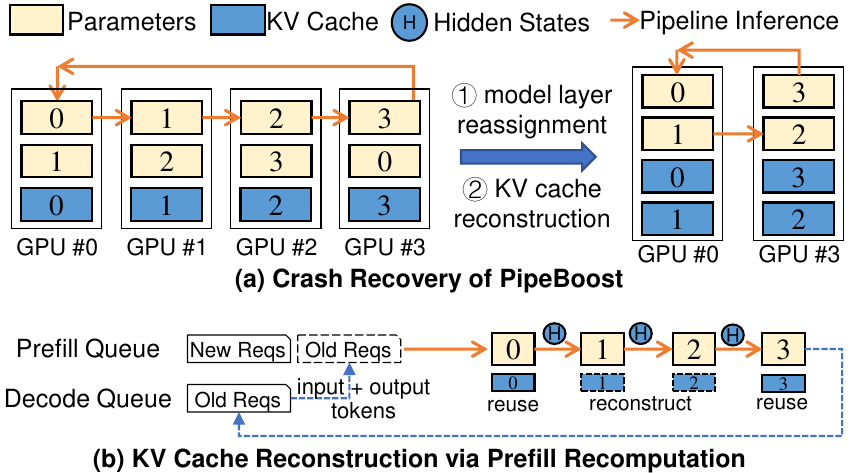} 
  \caption{\textbf{PipeBoost Recovery.}}
  \label{fig:ftpp_recovery} 
\end{figure}

\subsubsection{Pipeline-Parallel Recovery}
PipeBoost introduces rapid recovery techniques to address the issues in \S\ref{sec:need_for_fault_tolerance}.
The recovery process consists of two steps. 
The first step is model layer reassignment, which quickly reconstructs an efficient inference pipeline. 
The second step is KV cache reconstruction, enabling fast continuation of the inference process by reusing existing results.
We assume that all GPUs on a GPU server are initialized, as PipeBoost is designed to utilize the parallel bandwidth and computation capability of multiple GPUs.
Crashes may occur during either the model loading stage (\S\ref{sec:pp_model_loading}) or the model inference stage (\S\ref{sec:pp_model_inference}). 
This section discusses both scenarios separately.

\noindent
\textbf{Recovery for model loading.}
When an LLM instance crash is detected during model loading, PipeBoost dynamically adjusts the number of model layers loaded by the remaining operational instances. 
The adjustment follows two principles:
\begin{itemize}[noitemsep,topsep=0pt,parsep=0pt,partopsep=0pt]
    \item \textbf{Load Balance}: the workload among GPUs should be evenly distributed. 
    This paper focuses on scenarios where homogeneous GPUs are deployed on a single GPU server, meaning that assigning the same number of model layers to each GPU ensures balanced load. 
    For heterogeneous GPUs, a more sophisticated model partitioning strategy is required.
    \item \textbf{Layer Contiguity}: GPUs are assigned adjacent model layers whenever possible. 
    PipeBoost leverages pipeline parallelism during inference (\S\ref{sec:pp_model_inference}), so placing consecutive model layers on the same GPU minimizes the need for frequent inter-GPU synchronization.
\end{itemize}

Assume there are four GPUs, and the model is partitioned into 4 segments (Figure~\ref{fig:ftpp_recovery}a). 
GPU 0 sequentially loads model segments 0, 1, 2 and 3, while GPU 3 loads 3, 0, 1 and 2. GPUs 1 and 2 follow a similar pattern. 
Under normal circumstances, GPU 0 and GPU 3 only need to complete loading segments 0 and 3, respectively, to begin inference. 
Now, suppose GPU 1 and GPU 2 crash during model loading.
In this case, GPU 0's loading sequence remains unchanged, while PipeBoost adjusts GPU 3's loading order to 2, 3, 0 and 1. 
Once GPU 0 finishes loading 0 and 1, and GPU 3 completes 2 and 3, PipeBoost can proceed with inference execution.

\noindent
\textbf{Recovery for model inference.}
When an LLM instance crash is detected during model inference, PipeBoost first scans the GPUs to assess the distribution of loaded model layers and identifies a viable chain for pipeline parallelism to continue inference. 
If no such chain is found,  PipeBoost initiates the aforementioned model loading recovery process.

After model loading recovery, PipeBoost reconstructs the KV cache of the missing model layers. 
Specifically, PipeBoost distributes requests into two queues according to their stages (Figure~\ref{fig:ftpp_recovery}b). 
It inserts requests from the decode queue into the prefill queue to rebuild the KV cache. The input and output tokens processed so far are merged into a single input sequence and fed into the first model layer. 
For layers with an existing KV cache, PipeBoost recomputes the Q matrix and reuses the KV cache to generate hidden states for the next layer. 
For layers without a KV cache, PipeBoost performs a full prefill operation and stores the KV cache in GPU HBM. 
Once reconstruction is complete, the request is placed back into the decode queue.
\section{Evaluation}
\label{sec:figs}
PipeBoost is built upon Transformers~\cite{wolf-etal-2020-transformers} and PyTorch~\cite{paszkePyTorchImperativeStyle2019}. We adapt Transformers to enable PipeBoost to support LLMs such as OPT~\cite{zhangOPTOpenPretrained2022}, Falcon~\cite{almazroueiFalconSeriesOpen2023}, and Mistral~\cite{jiangMistral7B2023}, requiring only approximately 300 lines of code (LOC) modification for each model. This modular approach allows easy integration of additional models. Built on top of PyTorch, PipeBoost implements FTPP and incorporates continuous batching techniques~\cite{yuOrcaDistributedServing2022} to improve efficiency. Instead of using Ray~\cite{moritzRayDistributedFramework2018}, PipeBoost employs threads (via Python's \texttt{concurrent.futures}) to coordinate LLM instances within a GPU server, avoiding the significant initialization overhead associated with Ray.

We evaluate PipeBoost’s performance against state-of-the-art serverless LLM serving systems, focusing on low-latency startup.
We first introduce the evaluation setup (\S\ref{sec:eval_setup}), then analyze the startup process of base LLMs (\S\ref{sec:eval_starup}) and LoRA-based LLMs (\S\ref{sec:eval_lora}), followed by assessing  scalability across varying GPU counts, batch sizes, prompt lengths and improvements from epoch-based adapter switching (\S\ref{sec:eval_scalability}), ending with evaluating fault-tolerance mechanisms (\S\ref{sec:eval_recovery}). 

\subsection{Evaluation Setup}\label{sec:eval_setup}

\textbf{Device Setup.} We conduct our experiments on two servers, each equipped with 512GB DDR4 memory, an Intel Xeon Platinum 8338C CPU, and a Samsung 980 PRO 2TB NVMe SSD. One server is equipped with four NVIDIA RTX 4090 GPUs (24~GB memory), while the other contains two NVIDIA A100 GPUs (40~GB memory). Unless explicitly noted, all evaluations are performed on A100 GPUs.

\noindent\textbf{Model and Datasets.} We utilize advanced and widely-studied LLMs in system research including Mistral, OPT and Falcon, with parameter sizes of 1.3B, 2.7B, 6.7B, 13B, and 30B. 
We use the GSM8K datasets \cite{cobbeTrainingVerifiersSolve2021}, including real user questions across varying lengths and domains, with dynamic response lengths, making it representative of practical workloads.

\noindent\textbf{Baseline Systems.} We evaluate PipeBoost by comparing it to two baselines: the widely used Transformers framework and an optimized implementation of ServerlessLLM \cite{fuServerlessLLMLowLatencyServerless2024}.
For a fair comparison, the model loading and inference components of ServerlessLLM are included, while Ray \cite{moritzRayDistributedFramework2018}, a cross-server scheduling component, is excluded. This exclusion is justified as Ray's initialization takes 4-6 seconds, which is time-intensive and beyond the scope of our study.

\noindent\textbf{Metrics.} 
To assess the startup latency for LLM serving, we measure \textbf{\textit{Time to First Token (TTFT)}}, defined as the time elapsed from receiving an inference request to generating the first output token.
Notably, TTFT is also a key metric in existing LLM serving systems that mitigate cold starts by preloading LLM parameters into GPU memory.
In our experiments, TTFT encompasses the time required for model loading.
Additionally, we use \textbf{\textit{request completion latency (referred to as latency)}} to assess the total time required to complete a request. 
We measure \textbf{\textit{recovery time}}, defined as the duration from the moment of a crash to the point where the system resumes service, to evaluate the system's fault-tolerance capabilities in recovering from failures.

\subsection{Startup Performance of Base LLM}\label{sec:eval_starup}

\begin{figure}[t]
\centering
\includegraphics{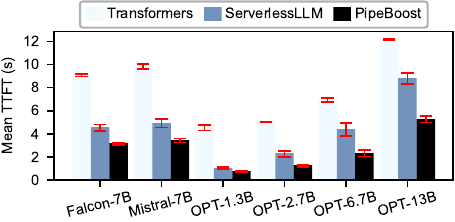}
\caption{\textbf{Mean TTFT across various base LLMs.}}
\label{fig:latency}
\end{figure}

\begin{figure}[t]
\centering
\includegraphics{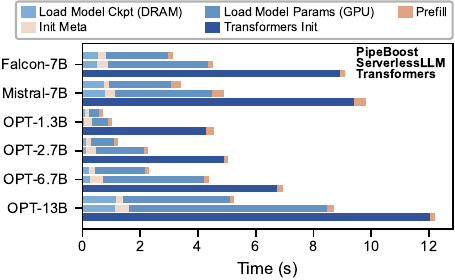}
\caption{\textbf{Mean TTFT breakdown.} Load Model Ckpt: loading model checkpoints from SSD to DRAM, Init Meta: initializing model metadata, Load Model Params: transferring checkpoints to GPUs and loading into model parameters, Prefill: performing prefill operations. Notably, Transformers follows its own loading procedure (Transformers Init).}
\label{fig:latency_breakdown}
\end{figure}

To evaluate PipeBoost's startup performance, we measure TTFT across various model architectures and sizes. 
The test launches one model on each GPU, simulating user input with a prompt size of 64 and a batch size of 64. 
The evaluation begins by loading the model checkpoint from an NVMe SSD and ends when the first token is output. 
As the test involves cold starts, our TTFT results may be higher compared to works focused on warm starts or pre-loaded models.

\noindent\textbf{Starup Time.}
We first compare the overall startup performance of PipeBoost with the baselines, as shown in Figure \ref{fig:latency}. The evaluation includes Falcon-7B, Mistral-7B, and the OPT series (OPT-1.3B, OPT-2.7B, OPT-6.7B, and OPT-13B). PipeBoost consistently outperforms both Transformers and ServerlessLLM, achieving TTFT reductions of 30\% to 85\% across all tested models. For example, the TTFT for Mistral-7B is reduced by 65\% compared to Transformers and 30\% compared to ServerlessLLM. These results demonstrate the efficiency of PipeBoost’s pipeline-parallel model loading.

Furthermore, PipeBoost shows significant advantages in handling larger models. For example, the TTFT for OPT-1.3B is reduced by 30.2\% compared to ServerlessLLM, while for OPT-13B, the reduction reaches 40.3\%. These reductions are attributed to PipeBoost’s ability to incrementally load model layers across multiple GPUs, minimizing idle GPU time and overlapping loading with inference.

\noindent\textbf{Startup Time Breakdown.}
To analyze the sources of startup latency, we decompose the startup process into key stages, following the order in which they occur during model startup. 
Figure \ref{fig:latency_breakdown} illustrates the contributions of each stage to the overall startup time for various model types, comparing PipeBoost with Transformers and ServerlessLLM. The results show that checkpoint loading (ranging from 9.05\% for OPT-2.7B to 22.1\% for OPT-13B) and parameter loading (ranging from 45.6\% for OPT-1.3B to 70.7\% for OPT-13B) are the most time-consuming stages during startup. 
For instance, in the case of OPT-13B, these two stages take 7.9 seconds with ServerlessLLM, compared to 4.8 seconds with PipeBoost, accounting for 92.8\% and 91.9\% of the overall startup latency.

PipeBoost significantly shortens these two stages, leading to a faster overall startup. For instance, with the Mistral-7B model, it reduces the time for these stages from 4.1 seconds in ServerlessLLM to 2.8 seconds, contributing to a total startup latency reduction from 4.9 seconds to 3.4 seconds.

\subsection{Startup Performance of LoRA LLM}\label{sec:eval_lora}

\begin{figure}[t]
\centering
\includegraphics{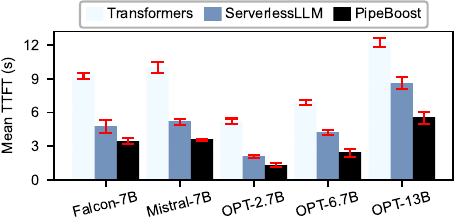}
\caption{\textbf{Mean TTFT across various LoRA LLMs.}}
\label{fig:latency_with_lora}
\end{figure}

\begin{table}[ht]
\centering
\caption{\textbf{Startup Time Breakdown with LoRA}}
\label{tab:latency_with_lora_breakdown}
\large
\resizebox{\columnwidth}{!}{%
\begin{tabular}{@{}lccc@{}}
\toprule
\multirow{2}{*}{\textbf{Stage}}     & \textbf{Mistral-7B}              & \textbf{OPT-13B}                  \\ 
                                    & \textbf{Time(s) / Rate(\%)}      & \textbf{Time(s) / Rate(\%)}       \\ \midrule
                                 
\multicolumn{3}{c}{PipeBoost} \\
\cmidrule(l{0.2cm}r{0.2cm}){1-3}

Load Model Ckpts (DRAM)          & 1.105 / 30.75                    & 1.157 / 20.86                     \\
Load Lora Ckpt (DRAM)            & 0.006 / \hphantom{0}0.17         & 0.010 / \hphantom{0}0.18           \\
Init Meta                        & 0.158 / \hphantom{0}4.40         & 0.229 / \hphantom{0}4.13          \\
Load Model Params (GPU)          & 1.970 / 54.81                    & 3.989 / 71.91                     \\
Load Lora Params (GPU)           & 0.025 / \hphantom{0}0.70         & 0.021 / \hphantom{0}0.38          \\
Prefill                          & 0.330 / \hphantom{0}9.18         & 0.141 / \hphantom{0}2.54          \\
\textbf{Total}                   & \textbf{3.594 / 100.0}             & \textbf{5.547 / 100.0}              \\

\cmidrule(){1-3}
\multicolumn{3}{c}{ServerlessLLM} \\ 
\cmidrule(l{0.2cm}r{0.2cm}){1-3}

Load Model Ckpts (DRAM)          & 1.091 / 20.92                         & 1.167 / 13.54                        \\
Load Lora Ckpt (DRAM)            & 0.005 / \hphantom{0}0.10              & 0.006 / \hphantom{0}0.07             \\
Init Meta                        & 0.267 / \hphantom{0}5.12              & 0.425 / \hphantom{0}4.93             \\
Load Model Params (GPU)          & 3.468 / 66.49                         & 6.815 / 79.08                        \\
Load Lora Params (GPU)           & 0.053 / \hphantom{0}1.02              & 0.041 / \hphantom{0}0.48             \\
Prefill                          & 0.332 / \hphantom{0}6.37              & 0.164 / \hphantom{0}1.90             \\
\textbf{Total}                   & \textbf{5.216 / 100.0}                & \textbf{8.618 / 100.0}               \\ \bottomrule

\end{tabular}%
}
\end{table}

LoRA, a parameter-efficient fine-tuning technique, is widely adopted for LLMs. PipeBoost is designed to efficiently support LoRA-enhanced models. In this section, we evaluate PipeBoost's startup latency for serving LoRA-equipped LLMs. The experimental setup mirrors \S\ref{sec:eval_starup}. 
The key difference is that each model starts with one LoRA adapter. This consistent setup ensures comparability while specifically focusing on the performance of LoRA-enabled scenarios.

\noindent\textbf{Startup Time.}
To assess the impact of LoRA on startup performance, we measure the TTFT for LoRA-based LLMs such as Mistral-7B, Falcon-7B, OPT-6.7B, and OPT-13B. 
As shown in Figure \ref{fig:latency_with_lora}, PipeBoost maintains its advantage over both Transformers and ServerlessLLM in LoRA-based LLM.
The reasons for PipeBoost's superior performance are similar to those observed in the base model evaluation (\S\ref{sec:eval_starup}), where pipeline-parallel model loading and inference play a crucial role in reducing latency.

A comparison of the performance between base LLMs (Figure~\ref{fig:latency}) and LoRA-enhanced LLMs (Figure~\ref{fig:latency_with_lora}) reveals that PipeBoost introduces minimal latency overhead in supporting LoRA. For example, with Mistral-7B, PipeBoost exhibits a TTFT increase of only 0.18 seconds, approximately a 5\% increase compared to the non-LoRA configuration. These results highlight PipeBoost’s ability to seamlessly integrate LoRA while preserving its pipeline-parallel efficiency.

\noindent\textbf{Startup Time Breakdown.}
Table \ref{tab:latency_with_lora_breakdown} provide a breakdown of the startup process for LoRA LLMs, illustrating how LoRA impacts startup time. 
When LoRA is enabled, the checkpoint loading stage involves additional steps to load the LoRA adapters into main memory, which are subsequently transferred to the GPU. 
Results show that additional steps from LoRA add negligible overhead to the total startup latency. 
Specifically, for Mistral-7B, a single LoRA adapter introduces only a 0.87\% overhead. For OPT-13B, the overhead is just 0.56\%. 
PipeBoost's pipeline-parallel model loading technique further reduces the loading time of LoRA adapters by approximately \textit{N} times, where \textit{N} represents the number of GPUs concurrently involved in the loading process.

The results here also demonstrate PipeBoost's ability to efficiently infer with multiple LoRA adapters. 
Since the primary overhead remains in the base model, PipeBoost can load dozens of (and even more) LoRA adapters into GPU memory with minimal cost. 
Combined with the epoch-based adapter switching technique (which will be evaluated in \S\ref{sec:eval_scalability}), PipeBoost is capable of efficiently supporting inference with multiple LoRA adapters based on the same base model.

\subsection{Scalability Analysis}\label{sec:eval_scalability}

\begin{figure}[t]
\centering
\includegraphics{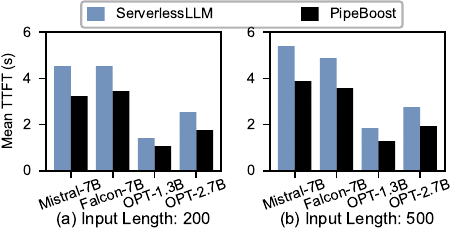}
\caption{\textbf{Mean TTFT with varying input lengths.}}
\label{fig:scala_input_len}
\end{figure}

\begin{figure}[t]
\centering
\includegraphics{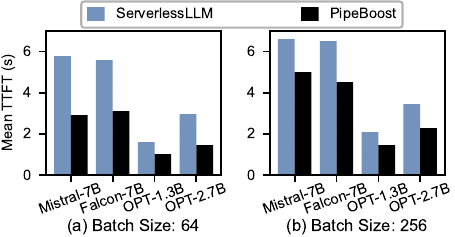}
\caption{\textbf{Mean TTFT with varying batch sizes.}}
\label{fig:scala_batch_size}
\end{figure}

\begin{figure}[t]
\centering
    \begin{minipage}[t]{0.21\textwidth} 
        \centering
        \includegraphics[width=\textwidth]{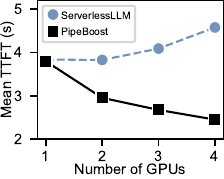}
        \caption{\textbf{Mean TTFT with varying GPU counts.} Lower is better.}
        \label{fig:split_influence}
    \end{minipage}
    \hspace{8pt}
    \begin{minipage}[t]{0.21\textwidth} 
        \centering
        \includegraphics[width=\textwidth]{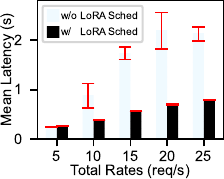}
        \caption{\textbf{Effect of epoch-based adapter switching technique.}}
        \label{fig:epoch_base_adapter_switching}
    \end{minipage}
\setlength{\abovecaptionskip}{0pt}
\setlength{\belowcaptionskip}{0pt}
\end{figure}

This section analyzes the scalability of PipeBoost under varying workload conditions, including input length, batch size, the number of GPUs, and the impact of epoch-based adapter switching for LoRA LLMs. This analysis demonstrates PipeBoost's ability to handle dynamic workloads efficiently and scale effectively across hardware resources.

\noindent\textbf{Scalability in Input Lenght and Batch Size.}
PipeBoost’s inference performance is further evaluated under varying input lengths and batch sizes to assess its adaptability to diverse workload scenarios. As shown in Figure \ref{fig:scala_input_len} and \ref{fig:scala_batch_size}, PipeBoost consistently outperforms ServerlessLLM in terms of TTFT across all input length and batch size configurations.

Figure \ref{fig:scala_input_len} illustrates the effect of input lengths (200 and 500 tokens) on TTFT. PipeBoost achieves significantly lower TTFT compared to ServerlessLLM across all tested models. For instance, with an input length of 200 tokens, PipeBoost reduces TTFT for Mistral-7B from 4.5 seconds to 3.2 seconds, a 28.9\% reduction compared to ServerlessLLM. Similarly, for an input length of 500 tokens, PipeBoost reduces TTFT for Mistral-7B from 5.4 seconds to 3.9 seconds, a 27.8\% reduction compared to ServerlessLLM. 

Figure \ref{fig:scala_batch_size} illustrates the influence of batch sizes (64 and 256) on TTFT. PipeBoost effectively handles larger batch sizes with minimal degradation in TTFT, showcasing its scalability for high-throughput workloads. For a batch size of 64, PipeBoost reduces TTFT for Falcon-7B from 5.6 seconds to 3.1 seconds, a 44.6\% reduction compared to ServerlessLLM. Even with a batch size of 256, PipeBoost maintains its advantage, reducing TTFT for Falcon-7B from 6.5 seconds to 4.5 seconds, a 30.8\% reduction. 

However, with the increase of input length and batch size, the performance gap between PipeBoost and ServerlessLLM narrows slightly, possibly due to the increased inter-GPU communication overhead of pipeline parallelism.

\noindent\textbf{Scalability with Multiple GPUs.}
To assess the scalability of PipeBoost with increasing GPU counts, we further measure the TTFT for Mistral-7B, building on the experimental settings outlined in \S\ref{sec:eval_starup}. The experiment is conducted on a server equipped with 4 NVIDIA RTX 4090 GPUs, with each GPU hosting one model instance. 

As shown in Figure \ref{fig:split_influence}, ServerlessLLM experiences a latency increase of up to 19.3\% as more GPUs are added. In contrast, PipeBoost achieves consistent TTFT reductions, ranging from 22.8\% to 46.5\%, compared to ServerlessLLM as the number of GPUs increases. Notably, with 4 GPUs, PipeBoost reduces TTFT by 35.5\% compared to the single-GPU configuration. The results demonstrate PipeBoost’s ability to scale, with the potential for further reductions in startup latency as GPU resources continue to grow.

\noindent\textbf{Effect of epoch-based LoRA adapter switching technique.}
To evaluate the effectiveness of PipeBoost’s epoch-based adapter switching mechanism, we conducted experiments with two LoRA-equipped LLMs on the GSM8K dataset.
The experiment gradually increases request rates. Dynamic workloads were simulated by switching LoRA adapters for user prompts with a switching probability of 20\%.

We measure the completion latency to include the evaluation of decode stages. 
The epoch-based adapter switching technique performs batching and scheduling on requests for different LoRA adapters (details in \S\ref{sec:design_pp_inference_lora}).
The baseline system serves requests as they arrive without LoRA scheduling, performing adapter switching eagerly.

As shown in Figure \ref{fig:epoch_base_adapter_switching}, PipeBoost (w/ LoRA Sched) significantly outperforms the baseline (w/o LoRA Sched). Without scheduling, mean latency increases from 0.3 seconds to 2.1 seconds as the request rate rises, with variance reaching 0.4 seconds at 20 RPS (requests per second). In comparison, epoch-based switching reduces mean latency by 63.1\% at 25 RPS and maintains much lower variance, peaking at nearly 0.8 microseconds at 5 RPS. 

\subsection{Fault Tolerance and Robustness}\label{sec:eval_recovery}
\begin{figure}[t]
\centering
\includegraphics{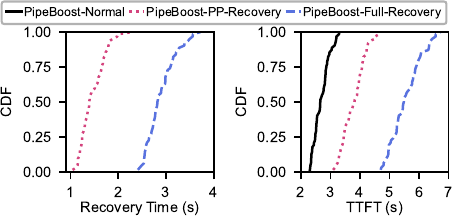}
\caption{\textbf{Recovery time and TTFT of crash recovery for the model loading stage.}}
\label{fig:load_recover_cdf}
\end{figure}

In this section, we evaluate the fault tolerance of PipeBoost during both the model loading and inference stages. 
Fault tolerance is crucial for maintaining system reliability in large-scale deployments where GPU failures are common.

We perform evaluations with Mistral-7B under three configurations: (a) LLM startup with no crashes (PipeBoost-Normal), (b) crash recovery with Pipeline-Parallel Recovery (PipeBoost-PP-Recovery), and (c) full recovery involving a complete restart of the pipeline-parallel loading and inference processes (PipeBoost-Full-Recovery). 

\noindent\textbf{Recovery for model loading.}
PipeBoost's recovery performance for the model loading stage is evaluated on a server with 4 NVIDIA RTX 4090 GPUs. To simulate realistic failure scenarios, errors were introduced in 2 GPUs during the loading process to trigger the recovery mechanism. 
Figure \ref{fig:load_recover_cdf} presents the recovery time and TTFT distributions.

PipeBoost-PP-Recovery demonstrates significant advantages over PipeBoost-Full-Recovery. For recovery time, PipeBoost-PP-Recovery achieves a median of 1.39 seconds, compared to 2.81 seconds for PipeBoost-Full-Recovery, yielding a 50.5\% latency reduction. In terms of TTFT, PipeBoost-PP-Recovery reduces the median time to 3.8 seconds, compared to 5.5 seconds for PipeBoost-Full-Recovery, closely approaching the PipeBoost-Normal baseline of 2.66 seconds.

PipeBoost minimizes disruption and ensures service continuity, by quickly reconstructing pipeline inference chains and redistributing workloads across operational GPUs.

\begin{figure}[t]
\centering
\includegraphics{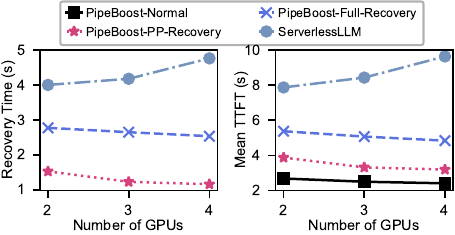}
\caption{\textbf{Impact of GPU Count on PipeBoost’s Crash Recovery.}}
\label{fig:load_recover_card_num}
\end{figure}

\noindent\textbf{Impact of GPU counts on PipeBoost’s crash recovery.}
We have previously demonstrated that startup performance improves with an increasing number of GPUs (Figure \ref{fig:split_influence}). Here, we further investigate whether recovery can similarly gain benefits. Figure \ref{fig:load_recover_card_num} shows the TTFT results for scenarios involving 2, 3, and 4 GPUs with a single GPU failure.

The results indicate that as the number of GPUs increases, the system is able to recover more quickly due to the additional resources available for redistributing workloads. For example, with 2 GPUs, the TTFT for recovery is 3.9 seconds, which improves to 3.3 seconds with 3 GPUs and further to 3.2 seconds with 4 GPUs, representing a 17.9\% reduction from 2 to 4 GPUs. 
However, the improvement slows between 3 and 4 GPUs, possibly due to increased communication and synchronization overhead between GPUs.

\begin{figure}[t]
\centering
\includegraphics{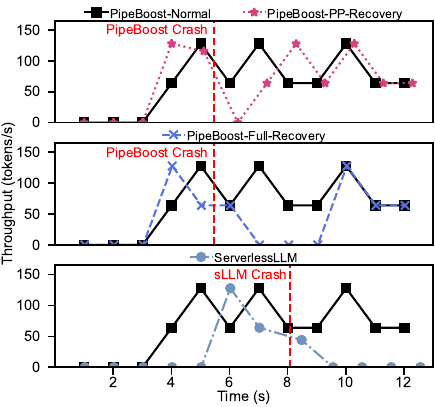}
\caption{\textbf{System throughput during crash recovery for the model inference stage.}}
\label{fig:infer_recover}
\end{figure}

\noindent\textbf{Recovery for model inference.}
In addition to evaluating recovery during model loading, it is essential to assess the impact of GPU crashes during inference. Once PipeBoost successfully enters the pipeline-parallel inference stage, we simulate GPU failures by inducing errors during the inference chain and observe the behavior of the Pipeline-Parallel Recovery mechanism. Figure \ref{fig:infer_recover} shows the throughput (in tokens per second) over time for PipeBoost.

When a GPU crash occurs during inference (at approximately 6 seconds), the throughput of PipeBoost-PP-Recovery reduces from 128 tokens/s to 64 tokens/s but recovers to a normal state within 2 seconds. 
However, PipeBoost-Full-Recovery experiences a complete halt in throughput for nearly 4 seconds before resuming at a normal rate. 

We observe that ServerlessLLM is slower than PipeBoost not only during startup but also during recovery.
ServerlessLLM requires approximately 5 seconds to start, during which its throughput remains near zero. In contrast, PipeBoost's throughput begins increasing at the 3-second mark. Similarly, ServerlessLLM takes nearly 5 seconds to recover from failures, which is 2.5$\times$ longer than PipeBoost's recovery time.

We also observe that under both normal and crash recovery scenarios, PipeBoost’s throughput fluctuates. This fluctuation is primarily due to the differing computational complexities of prefill and decode operations, which is a normal phenomenon and does not affect the argument for PipeBoost's low-latency startup. For instance, at throughput low points, there are more prefill requests, while at high points, there are more decode requests. Since prefill is computationally slower than decode, it reduces overall throughput.

\section{Related Works}

\noindent
\textbf{LLM serving systems.}
Recent advancements in LLM serving \cite{yuOrcaDistributedServing2022, liAlpaServeStatisticalMultiplexing2023a,jeongFastEfficientModel2023,patelSplitwiseEfficientGenerative2024} primarily focus on optimizing inference performance, overlooking the model loading stage. 
However, several of these techniques can complement PipeBoost's pipeline-parallel model inference design. 
For example, Orca \cite{yuOrcaDistributedServing2022} introduces a continuous batching technique to enhance GPU utilization, which PipeBoost adopts. 
Splitwise \cite{patelSplitwiseEfficientGenerative2024} and DistServe \cite{zhongDistServeDisaggregatingPrefill2024} separate prefill and decode computations across different GPUs, a method that can be applied in PipeBoost when both GPU groups for prefill and decode computations complete loading an available pipeline inference chain.

Many works \cite{liAlpaServeStatisticalMultiplexing2023a,sunLlumnixDynamicScheduling2024a,agrawalTamingThroughputLatencyTradeoff2024,wuDLoRADynamicallyOrchestrating2024,wuLoongServeEfficientlyServing2024a} propose inference strategies that PipeBoost can adopt once all GPUs complete loading the full model. 
vLLM \cite{kwonEfficientMemoryManagement2023c} introduces paged attention to optimize GPU memory management. 
AlpaServe \cite{liAlpaServeStatisticalMultiplexing2023a} employs model parallelism to expedite serving large DNN models across GPU clusters. 
Llumnix \cite{sunLlumnixDynamicScheduling2024a} dynamically reschedules inference requests across LLM instances, enabling systems to adapt to unpredictable workload dynamics. 
Techniques like chunked-prefill and stall-free batching in Sarathi-Serve \cite{agrawalTamingThroughputLatencyTradeoff2024} strike a balance between throughput and latency.
dLoRA \cite{wuDLoRADynamicallyOrchestrating2024} orchestrates requests and LoRA adapters across replicas to achieve effective load balancing. 
LoongServe \cite{wuLoongServeEfficientlyServing2024a} introduces elastic sequence parallelism, offering adaptability to variable-length requests in different processing phases.

\noindent
\textbf{Optimizations for serverless cold-starts.}
The cold-start issue in serverless computing has been present since the early days of CPU-based computation.
In GPU computing, this problem becomes even more severe due to the increased complexity of GPU application initialization \cite{yangOndemandParallelCheckpoint2024}, further exacerbated by the rapid growth in LLM parameters \cite{fuServerlessLLMLowLatencyServerless2024}.
Numerous optimizations are proposed, including rapid image retrieval \cite{wangFaaSNetScalableFast2021}, streamlined isolation mechanisms \cite{liRunDLightweightSecure2022,oakesSOCKRapidTask2018}, snapshot-based restoration \cite{aoFaaSnapFaaSMade2022,caddenSEUSSSkipRedundant2020,duCatalyzerSubmillisecondStartup2020,shillakerFAASMLightweightIsolation2020a,ustiugovBenchmarkingAnalysisOptimization2021}, dynamic pre-provisioning of resources \cite{shahradServerlessWildCharacterizing2020}, and process fork \cite{akkusSANDHighperformanceServerless2018,weiNoProvisionedConcurrency2023b}. 
While these methods effectively minimize startup times for containers or virtual machines, they do not address the high loading latency of LLM parameters.

ServerlessLLM utilizes high-performance storage within GPU servers to access LLMs. 
PipeBoost further enhances this process through FTPP techniques.
To achieve the fastest cold start, ServerlessLLM estimates and compares model startup times and service migration durations across GPUs, selecting the most suitable server. In contrast, PipeBoost simplifies this process by initializing the same model on the same GPU server, avoiding the overhead of migration and estimation.
With advancements like GPUDirect Storage technology and NVMe arrays, the time required for GPUs to load LLM parameters from DRAM and NVMe SSDs is expected to converge. Under such conditions, ServerlessLLM's selection strategy and live migration approach may become less effective, while PipeBoost stands to gain performance improvements.

\noindent
\textbf{Pipeline parallelism.}
Pipeline parallelism is commonly used in model training \cite{narayananPipeDreamGeneralizedPipeline2019,huangGPipeEfficientTraining2019,fanDAPPLEPipelinedData2021,liTeraPipeTokenLevelPipeline2021,narayananMemoryEfficientPipelineParallelDNN2021}, but we apply the technique to the cold-start process of serverless LLMs, demonstrating its capability to rapidly initialize inference services.
Additionally, we show how to provide fault tolerance for pipeline parallelism, thereby accelerating the recovery process of LLM serving systems.
Moreover, we highlight the need to switch from pipeline parallelism to alternative parallel strategies when the request density on GPU servers exceeds a certain threshold to avoid excessive communication overhead.

\section{Conclusion}

We present PipeBoost, a low-latency inference engine for serverless LLM deployments, designed to achieve fast startup with robust fault-tolerant performance. 
PipeBoost significantly reduces cold-start times compared to state-of-the-art systems, supports LoRA-equipped LLMs with efficient inference enabled by its epoch-based adapter switching mechanism, and ensures minimal service disruption during GPU failures with its fault-tolerant recovery design. 
Extensive evaluations demonstrate PipeBoost's efficiency, scalability, and fault tolerance in multi-GPU environments.

\bibliographystyle{plain}
\bibliography{reference/reference}

\begin{thebibliography}{10}

\bibitem{AmazonSageMaker}
Amazon {{SageMaker}}.
\newblock \url{https://aws.amazon.com/sagemaker/}.

\bibitem{AzureFunctionsServerlessa}
Azure {{Functions}} -- {{Serverless Functions}} in {{Computing}}.
\newblock \url{https://azure.microsoft.com/en-us/products/functions}.

\bibitem{GitHubCopilotYoura}
{{GitHub Copilot}}.
\newblock \url{https://github.com/features/copilot}.

\bibitem{HuggingFaceGenerative}
Hugging {{Face Generative AI Services}} ({{HUGS}}).
\newblock \url{https://huggingface.co/docs/hugs/index}.

\bibitem{InformationBeautiful}
Information is {{Beautiful}}: the rise of generative-ai like chatgpt.
\newblock \url{https://informationisbeautiful.net/visualizations/}.

\bibitem{IntroducingChatGPT}
Introducing {{ChatGPT}}.
\newblock \url{https://openai.com/index/chatgpt/}.

\bibitem{IntroducingLlama31}
Introducing {{Llama}} 3.1: {{Our}} most capable models to date.
\newblock \url{https://ai.meta.com/blog/meta-llama-3-1/}.

\bibitem{IntroducingNewBing}
Introducing the new {{Bing}}.
\newblock \url{https://www.microsoft.com/en-us/edge/features/the-new-bing}.

\bibitem{agrawalTamingThroughputLatencyTradeoff2024}
Amey Agrawal, Nitin Kedia, Ashish Panwar, Jayashree Mohan, Nipun Kwatra, Bhargav Gulavani, Alexey Tumanov, and Ramachandran Ramjee.
\newblock Taming \{\vphantom\}{{Throughput-Latency}}\vphantom\{\} {{Tradeoff}} in \{\vphantom\}{{LLM}}\vphantom\{\} {{Inference}} with \{\vphantom\}{{Sarathi-Serve}}\vphantom\{\}.
\newblock In {\em 18th {{USENIX Symposium}} on {{Operating Systems Design}} and {{Implementation}} ({{OSDI}} 24)}, pages 117--134, 2024.

\bibitem{akkusSANDHighperformanceServerless2018}
Istemi~Ekin Akkus, Ruichuan Chen, Ivica Rimac, Manuel Stein, Klaus Satzke, Andre Beck, Paarijaat Aditya, and Volker Hilt.
\newblock {{SAND}}: Towards high-performance serverless computing.
\newblock In {\em Proceedings of the 2018 {{USENIX Conference}} on {{Usenix Annual Technical Conference}}}, {{USENIX ATC}} '18, pages 923--935, USA, July 2018. USENIX Association.

\bibitem{almazroueiFalconSeriesOpen2023}
Ebtesam Almazrouei, Hamza Alobeidli, Abdulaziz Alshamsi, Alessandro Cappelli, Ruxandra Cojocaru, M{\'e}rouane Debbah, {\'E}tienne Goffinet, Daniel Hesslow, Julien Launay, Quentin Malartic, Daniele Mazzotta, Badreddine Noune, Baptiste Pannier, and Guilherme Penedo.
\newblock The {{Falcon Series}} of {{Open Language Models}}, November 2023.

\bibitem{aoFaaSnapFaaSMade2022}
Lixiang Ao, George Porter, and Geoffrey~M. Voelker.
\newblock {{FaaSnap}}: {{FaaS}} made fast using snapshot-based {{VMs}}.
\newblock In {\em Proceedings of the {{Seventeenth European Conference}} on {{Computer Systems}}}, {{EuroSys}} '22, pages 730--746, New York, NY, USA, March 2022. Association for Computing Machinery.

\bibitem{brownLanguageModelsAre2020}
Tom~B. Brown, Benjamin Mann, Nick Ryder, Melanie Subbiah, Jared Kaplan, Prafulla Dhariwal, Arvind Neelakantan, Pranav Shyam, Girish Sastry, Amanda Askell, Sandhini Agarwal, Ariel {Herbert-Voss}, Gretchen Krueger, Tom Henighan, Rewon Child, Aditya Ramesh, Daniel~M. Ziegler, Jeffrey Wu, Clemens Winter, Christopher Hesse, Mark Chen, Eric Sigler, Mateusz Litwin, Scott Gray, Benjamin Chess, Jack Clark, Christopher Berner, Sam McCandlish, Alec Radford, Ilya Sutskever, and Dario Amodei.
\newblock Language models are few-shot learners.
\newblock In {\em Proceedings of the 34th {{International Conference}} on {{Neural Information Processing Systems}}}, {{NIPS}} '20, pages 1877--1901, Red Hook, NY, USA, December 2020. Curran Associates Inc.

\bibitem{caddenSEUSSSkipRedundant2020}
James Cadden, Thomas Unger, Yara Awad, Han Dong, Orran Krieger, and Jonathan Appavoo.
\newblock {{SEUSS}}: Skip redundant paths to make serverless fast.
\newblock In {\em Proceedings of the {{Fifteenth European Conference}} on {{Computer Systems}}}, {{EuroSys}} '20, pages 1--15, New York, NY, USA, April 2020. Association for Computing Machinery.

\bibitem{cobbeTrainingVerifiersSolve2021}
Karl Cobbe, Vineet Kosaraju, Mohammad Bavarian, Mark Chen, Heewoo Jun, Lukasz Kaiser, Matthias Plappert, Jerry Tworek, Jacob Hilton, Reiichiro Nakano, Christopher Hesse, and John Schulman.
\newblock Training {{Verifiers}} to {{Solve Math Word Problems}}, November 2021.

\bibitem{deepseek-aiDeepSeekV2StrongEconomical2024}
{DeepSeek-AI}.
\newblock {{DeepSeek-V2}}: {{A Strong}}, {{Economical}}, and {{Efficient Mixture-of-Experts Language Model}}, May 2024.

\bibitem{duCatalyzerSubmillisecondStartup2020}
Dong Du, Tianyi Yu, Yubin Xia, Binyu Zang, Guanglu Yan, Chenggang Qin, Qixuan Wu, and Haibo Chen.
\newblock Catalyzer: {{Sub-millisecond Startup}} for {{Serverless Computing}} with {{Initialization-less Booting}}.
\newblock In {\em Proceedings of the {{Twenty-Fifth International Conference}} on {{Architectural Support}} for {{Programming Languages}} and {{Operating Systems}}}, {{ASPLOS}} '20, pages 467--481, New York, NY, USA, March 2020. Association for Computing Machinery.

\bibitem{fanDAPPLEPipelinedData2021}
Shiqing Fan, Yi~Rong, Chen Meng, Zongyan Cao, Siyu Wang, Zhen Zheng, Chuan Wu, Guoping Long, Jun Yang, Lixue Xia, Lansong Diao, Xiaoyong Liu, and Wei Lin.
\newblock {{DAPPLE}}: A pipelined data parallel approach for training large models.
\newblock In {\em Proceedings of the 26th {{ACM SIGPLAN Symposium}} on {{Principles}} and {{Practice}} of {{Parallel Programming}}}, {{PPoPP}} '21, pages 431--445, New York, NY, USA, February 2021. Association for Computing Machinery.

\bibitem{fuServerlessLLMLowLatencyServerless2024}
Yao Fu, Leyang Xue, Yeqi Huang, Andrei-Octavian Brabete, Dmitrii Ustiugov, Yuvraj Patel, and Luo Mai.
\newblock \{\vphantom\}{{ServerlessLLM}}\vphantom\{\}: \{\vphantom\}{{Low-Latency}}\vphantom\{\} {{Serverless Inference}} for {{Large Language Models}}.
\newblock In {\em 18th {{USENIX Symposium}} on {{Operating Systems Design}} and {{Implementation}} ({{OSDI}} 24)}, pages 135--153, 2024.

\bibitem{gaoEmpiricalStudyLow2024}
Yanjie Gao, Yichen He, Xinze Li, Bo~Zhao, Haoxiang Lin, Yoyo Liang, Jing Zhong, Hongyu Zhang, Jingzhou Wang, Yonghua Zeng, Keli Gui, Jie Tong, and Mao Yang.
\newblock An {{Empirical Study}} on {{Low GPU Utilization}} of {{Deep Learning Jobs}}.
\newblock In {\em Proceedings of the {{IEEE}}/{{ACM}} 46th {{International Conference}} on {{Software Engineering}}}, {{ICSE}} '24, pages 1--13, New York, NY, USA, April 2024. Association for Computing Machinery.

\bibitem{grattafioriLlama3Herd2024}
Aaron Grattafiori, Abhimanyu Dubey, Abhinav Jauhri, Abhinav Pandey, Abhishek Kadian, Ahmad {Al-Dahle}, Aiesha Letman, Akhil Mathur, Alan Schelten, Alex Vaughan, Amy Yang, Angela Fan, Anirudh Goyal, Anthony Hartshorn, Aobo Yang, Archi Mitra, Archie Sravankumar, Artem Korenev, Arthur Hinsvark, Arun Rao, Aston Zhang, Aurelien Rodriguez, Austen Gregerson, Ava Spataru, Baptiste Roziere, Bethany Biron, Binh Tang, Bobbie Chern, Charlotte Caucheteux, Chaya Nayak, Chloe Bi, Chris Marra, Chris McConnell, Christian Keller, Christophe Touret, Chunyang Wu, Corinne Wong, Cristian~Canton Ferrer, Cyrus Nikolaidis, et~al.
\newblock The {{Llama}} 3 {{Herd}} of {{Models}}, November 2024.

\bibitem{huLoRALowRankAdaptation2021}
Edward~J. Hu, Yelong Shen, Phillip Wallis, Zeyuan {Allen-Zhu}, Yuanzhi Li, Shean Wang, Lu~Wang, and Weizhu Chen.
\newblock {{LoRA}}: {{Low-Rank Adaptation}} of {{Large Language Models}}, October 2021.

\bibitem{huCharacterizationLargeLanguage2024a}
Qinghao Hu, Zhisheng Ye, Zerui Wang, Guoteng Wang, Meng Zhang, Qiaoling Chen, Peng Sun, Dahua Lin, Xiaolin Wang, Yingwei Luo, Yonggang Wen, and Tianwei Zhang.
\newblock Characterization of {{Large Language Model Development}} in the {{Datacenter}}.
\newblock In {\em 21st {{USENIX Symposium}} on {{Networked Systems Design}} and {{Implementation}} ({{NSDI}} 24)}, pages 709--729, 2024.

\bibitem{huangENOVAAutoscalingCosteffective2024}
Tao Huang, Pengfei Chen, Kyoka Gong, Jocky Hawk, Zachary Bright, Wenxin Xie, Kecheng Huang, and Zhi Ji.
\newblock {{ENOVA}}: {{Autoscaling}} towards {{Cost-effective}} and {{Stable Serverless LLM Serving}}, May 2024.

\bibitem{huangGPipeEfficientTraining2019}
Yanping Huang, Youlong Cheng, Ankur Bapna, Orhan Firat, Mia~Xu Chen, Dehao Chen, HyoukJoong Lee, Jiquan Ngiam, Quoc~V. Le, Yonghui Wu, and Zhifeng Chen.
\newblock {{GPipe}}: Efficient training of giant neural networks using pipeline parallelism.
\newblock In {\em Proceedings of the 33rd {{International Conference}} on {{Neural Information Processing Systems}}}, number~10, pages 103--112. Curran Associates Inc., Red Hook, NY, USA, December 2019.

\bibitem{jeongFastEfficientModel2023}
Jinwoo Jeong, Seungsu Baek, and Jeongseob Ahn.
\newblock Fast and {{Efficient Model Serving Using Multi-GPUs}} with {{Direct-Host-Access}}.
\newblock In {\em Proceedings of the {{Eighteenth European Conference}} on {{Computer Systems}}}, {{EuroSys}} '23, pages 249--265, New York, NY, USA, May 2023. Association for Computing Machinery.

\bibitem{jiangMistral7B2023}
Albert~Q. Jiang, Alexandre Sablayrolles, Arthur Mensch, Chris Bamford, Devendra~Singh Chaplot, Diego de~las Casas, Florian Bressand, Gianna Lengyel, Guillaume Lample, Lucile Saulnier, L{\'e}lio~Renard Lavaud, Marie-Anne Lachaux, Pierre Stock, Teven~Le Scao, Thibaut Lavril, Thomas Wang, Timoth{\'e}e Lacroix, and William~El Sayed.
\newblock Mistral {{7B}}, October 2023.

\bibitem{jiangMegaScaleScalingLarge}
Ziheng Jiang, Haibin Lin, Yinmin Zhong, Qi~Huang, Yangrui Chen, Zhi Zhang, Yanghua Peng, Xiang Li, Cong Xie, Shibiao Nong, Yulu Jia, Sun He, Hongmin Chen, Zhihao Bai, Qi~Hou, Shipeng Yan, Ding Zhou, Yiyao Sheng, Zhuo Jiang, Haohan Xu, Haoran Wei, Zhang Zhang, Pengfei Nie, Leqi Zou, Sida Zhao, Liang Xiang, Zherui Liu, Zhe Li, Xiaoying Jia, Jianxi Ye, and Xin Jin.
\newblock {{MegaScale}}: {{Scaling Large Language Model Training}} to {{More Than}} 10,000 {{GPUs}}.

\bibitem{kwonEfficientMemoryManagement2023c}
Woosuk Kwon, Zhuohan Li, Siyuan Zhuang, Ying Sheng, Lianmin Zheng, Cody~Hao Yu, Joseph Gonzalez, Hao Zhang, and Ion Stoica.
\newblock Efficient {{Memory Management}} for {{Large Language Model Serving}} with {{PagedAttention}}.
\newblock In {\em Proceedings of the 29th {{Symposium}} on {{Operating Systems Principles}}}, {{SOSP}} '23, pages 611--626, New York, NY, USA, October 2023. Association for Computing Machinery.

\bibitem{liAlpaServeStatisticalMultiplexing2023a}
Zhuohan Li, Lianmin Zheng, Yinmin Zhong, Vincent Liu, Ying Sheng, Xin Jin, Yanping Huang, Zhifeng Chen, Hao Zhang, Joseph~E. Gonzalez, and Ion Stoica.
\newblock \{\vphantom\}{{AlpaServe}}\vphantom\{\}: {{Statistical Multiplexing}} with {{Model Parallelism}} for {{Deep Learning Serving}}.
\newblock In {\em 17th {{USENIX Symposium}} on {{Operating Systems Design}} and {{Implementation}} ({{OSDI}} 23)}, pages 663--679, 2023.

\bibitem{liTeraPipeTokenLevelPipeline2021}
Zhuohan Li, Siyuan Zhuang, Shiyuan Guo, Danyang Zhuo, Hao Zhang, Dawn Song, and Ion Stoica.
\newblock {{TeraPipe}}: {{Token-Level Pipeline Parallelism}} for {{Training Large-Scale Language Models}}, September 2021.

\bibitem{liRunDLightweightSecure2022}
Zijun Li, Jiagan Cheng, Quan Chen, Eryu Guan, Zizheng Bian, Yi~Tao, Bin Zha, Qiang Wang, Weidong Han, and Minyi Guo.
\newblock \{\vphantom\}{{RunD}}\vphantom\{\}: {{A Lightweight Secure Container Runtime}} for {{High-density Deployment}} and {{High-concurrency Startup}} in {{Serverless Computing}}.
\newblock In {\em 2022 {{USENIX Annual Technical Conference}} ({{USENIX ATC}} 22)}, pages 53--68, 2022.

\bibitem{lianUniversalCheckpointingEfficient2024}
Xinyu Lian, Sam~Ade Jacobs, Lev Kurilenko, Masahiro Tanaka, Stas Bekman, Olatunji Ruwase, and Minjia Zhang.
\newblock Universal {{Checkpointing}}: {{Efficient}} and {{Flexible Checkpointing}} for {{Large Scale Distributed Training}}.
\newblock {\em CoRR}, January 2024.

\bibitem{mohanAgileColdStarts2019}
Anup Mohan, Harshad Sane, Kshitij Doshi, Saikrishna Edupuganti, Naren Nayak, and Vadim Sukhomlinov.
\newblock Agile {{Cold Starts}} for {{Scalable Serverless}}.
\newblock In {\em 11th {{USENIX Workshop}} on {{Hot Topics}} in {{Cloud Computing}} ({{HotCloud}} 19)}, 2019.

\bibitem{mohanCheckFreqFrequentFineGrained2021}
Jayashree Mohan, Amar Phanishayee, and Vijay Chidambaram.
\newblock \{\vphantom\}{{CheckFreq}}\vphantom\{\}: {{Frequent}}, \{\vphantom\}{{Fine-Grained}}\vphantom\{\} \{\vphantom\}{{DNN}}\vphantom\{\} {{Checkpointing}}.
\newblock In {\em 19th {{USENIX Conference}} on {{File}} and {{Storage Technologies}} ({{FAST}} 21)}, pages 203--216, 2021.

\bibitem{moritzRayDistributedFramework2018}
Philipp Moritz, Robert Nishihara, Stephanie Wang, Alexey Tumanov, Richard Liaw, Eric Liang, Melih Elibol, Zongheng Yang, William Paul, Michael~I. Jordan, and Ion Stoica.
\newblock Ray: A distributed framework for emerging {{AI}} applications.
\newblock In {\em Proceedings of the 13th {{USENIX}} Conference on {{Operating Systems Design}} and {{Implementation}}}, {{OSDI}}'18, pages 561--577, USA, October 2018. USENIX Association.

\bibitem{narayananPipeDreamGeneralizedPipeline2019}
Deepak Narayanan, Aaron Harlap, Amar Phanishayee, Vivek Seshadri, Nikhil~R. Devanur, Gregory~R. Ganger, Phillip~B. Gibbons, and Matei Zaharia.
\newblock {{PipeDream}}: Generalized pipeline parallelism for {{DNN}} training.
\newblock In {\em Proceedings of the 27th {{ACM Symposium}} on {{Operating Systems Principles}}}, {{SOSP}} '19, pages 1--15, New York, NY, USA, October 2019. Association for Computing Machinery.

\bibitem{narayananMemoryEfficientPipelineParallelDNN2021}
Deepak Narayanan, Amar Phanishayee, Kaiyu Shi, Xie Chen, and Matei Zaharia.
\newblock Memory-{{Efficient Pipeline-Parallel DNN Training}}.
\newblock In {\em Proceedings of the 38th {{International Conference}} on {{Machine Learning}}}, pages 7937--7947. PMLR, July 2021.

\bibitem{oakesSOCKRapidTask2018}
Edward Oakes, Leon Yang, Dennis Zhou, Kevin Houck, Tyler Harter, Andrea~C. {Arpaci-Dusseau}, and Remzi~H. {Arpaci-Dusseau}.
\newblock {{SOCK}}: Rapid task provisioning with serverless-optimized containers.
\newblock In {\em Proceedings of the 2018 {{USENIX Conference}} on {{Usenix Annual Technical Conference}}}, {{USENIX ATC}} '18, pages 57--69, USA, July 2018. USENIX Association.

\bibitem{openaiGPT4TechnicalReport2024}
OpenAI, Josh Achiam, Steven Adler, Sandhini Agarwal, Lama Ahmad, Ilge Akkaya, Florencia~Leoni Aleman, Diogo Almeida, Janko Altenschmidt, Sam Altman, Shyamal Anadkat, Red Avila, Igor Babuschkin, Suchir Balaji, Valerie Balcom, Paul Baltescu, Haiming Bao, Mohammad Bavarian, Jeff Belgum, Irwan Bello, Jake Berdine, et~al.
\newblock {{GPT-4 Technical Report}}, March 2024.

\bibitem{paszkePyTorchImperativeStyle2019}
Adam Paszke, Sam Gross, Francisco Massa, Adam Lerer, James Bradbury, Gregory Chanan, Trevor Killeen, Zeming Lin, Natalia Gimelshein, Luca Antiga, Alban Desmaison, Andreas K{\"o}pf, Edward Yang, Zach DeVito, Martin Raison, Alykhan Tejani, Sasank Chilamkurthy, Benoit Steiner, Lu~Fang, Junjie Bai, and Soumith Chintala.
\newblock {{PyTorch}}: An imperative style, high-performance deep learning library.
\newblock In {\em Proceedings of the 33rd {{International Conference}} on {{Neural Information Processing Systems}}}, number 721, pages 8026--8037. Curran Associates Inc., Red Hook, NY, USA, December 2019.

\bibitem{patelSplitwiseEfficientGenerative2024}
Pratyush Patel, Esha Choukse, Chaojie Zhang, Aashaka Shah, {\'I}{\~n}igo Goiri, Saeed Maleki, and Ricardo Bianchini.
\newblock Splitwise: {{Efficient Generative LLM Inference Using Phase Splitting}}.
\newblock In {\em 2024 {{ACM}}/{{IEEE}} 51st {{Annual International Symposium}} on {{Computer Architecture}} ({{ISCA}})}, pages 118--132, June 2024.

\bibitem{qinMooncakeKVCachecentricDisaggregated2024a}
Ruoyu Qin, Zheming Li, Weiran He, Mingxing Zhang, Yongwei Wu, Weimin Zheng, and Xinran Xu.
\newblock Mooncake: {{A KVCache-centric Disaggregated Architecture}} for {{LLM Serving}}, July 2024.

\bibitem{rajbhandariZeROMemoryOptimizations2020}
Samyam Rajbhandari, Jeff Rasley, Olatunji Ruwase, and Yuxiong He.
\newblock {{ZeRO}}: Memory optimizations toward training trillion parameter models.
\newblock In {\em Proceedings of the {{International Conference}} for {{High Performance Computing}}, {{Networking}}, {{Storage}} and {{Analysis}}}, {{SC}} '20, pages 1--16, Atlanta, Georgia, November 2020. IEEE Press.

\bibitem{sahraeiXFaaSHyperscaleLow2023}
Alireza Sahraei, Soteris Demetriou, Amirali Sobhgol, Haoran Zhang, Abhigna Nagaraja, Neeraj Pathak, Girish Joshi, Carla Souza, Bo~Huang, Wyatt Cook, Andrii Golovei, Pradeep Venkat, Andrew Mcfague, Dimitrios Skarlatos, Vipul Patel, Ravinder Thind, Ernesto Gonzalez, Yun Jin, and Chunqiang Tang.
\newblock {{XFaaS}}: {{Hyperscale}} and {{Low Cost Serverless Functions}} at {{Meta}}.
\newblock In {\em Proceedings of the 29th {{Symposium}} on {{Operating Systems Principles}}}, {{SOSP}} '23, pages 231--246, New York, NY, USA, October 2023. Association for Computing Machinery.

\bibitem{shahradServerlessWildCharacterizing2020}
Mohammad Shahrad, Rodrigo Fonseca, {\'I}{\~n}igo Goiri, Gohar Chaudhry, Paul Batum, Jason Cooke, Eduardo Laureano, Colby Tresness, Mark Russinovich, and Ricardo Bianchini.
\newblock Serverless in the wild: Characterizing and optimizing the serverless workload at a large cloud provider.
\newblock In {\em Proceedings of the 2020 {{USENIX Conference}} on {{Usenix Annual Technical Conference}}}, {{USENIX ATC}}'20, pages 205--218, USA, July 2020. USENIX Association.

\bibitem{shillakerFAASMLightweightIsolation2020a}
Simon Shillaker and Peter Pietzuch.
\newblock {{FAASM}}: Lightweight isolation for efficient stateful serverless computing.
\newblock In {\em Proceedings of the 2020 {{USENIX Conference}} on {{Usenix Annual Technical Conference}}}, {{USENIX ATC}}'20, pages 419--433, USA, July 2020. USENIX Association.

\bibitem{sunLlumnixDynamicScheduling2024a}
Biao Sun, Ziming Huang, Hanyu Zhao, Wencong Xiao, Xinyi Zhang, Yong Li, and Wei Lin.
\newblock Llumnix: {{Dynamic Scheduling}} for {{Large Language Model Serving}}.
\newblock In {\em 18th {{USENIX Symposium}} on {{Operating Systems Design}} and {{Implementation}} ({{OSDI}} 24)}, pages 173--191, 2024.

\bibitem{thorpeBambooMakingPreemptible2023a}
John Thorpe, Pengzhan Zhao, Jonathan Eyolfson, Yifan Qiao, Zhihao Jia, Minjia Zhang, Ravi Netravali, and Guoqing~Harry Xu.
\newblock Bamboo: {{Making Preemptible Instances Resilient}} for {{Affordable Training}} of {{Large}} \{\vphantom\}{{DNNs}}\vphantom\{\}.
\newblock In {\em 20th {{USENIX Symposium}} on {{Networked Systems Design}} and {{Implementation}} ({{NSDI}} 23)}, pages 497--513, 2023.

\bibitem{ustiugovBenchmarkingAnalysisOptimization2021}
Dmitrii Ustiugov, Plamen Petrov, Marios Kogias, Edouard Bugnion, and Boris Grot.
\newblock Benchmarking, analysis, and optimization of serverless function snapshots.
\newblock In {\em Proceedings of the 26th {{ACM International Conference}} on {{Architectural Support}} for {{Programming Languages}} and {{Operating Systems}}}, {{ASPLOS}} '21, pages 559--572, New York, NY, USA, April 2021. Association for Computing Machinery.

\bibitem{wangFaaSNetScalableFast2021}
Ao~Wang, Shuai Chang, Huangshi Tian, Hongqi Wang, Haoran Yang, Huiba Li, Rui Du, and Yue Cheng.
\newblock \{\vphantom\}{{FaaSNet}}\vphantom\{\}: {{Scalable}} and {{Fast Provisioning}} of {{Custom Serverless Container Runtimes}} at {{Alibaba Cloud Function Compute}}.
\newblock In {\em 2021 {{USENIX Annual Technical Conference}} ({{USENIX ATC}} 21)}, pages 443--457, 2021.

\bibitem{wang2024boosting}
Shenzhi Wang, Chang Liu, Zilong Zheng, Siyuan Qi, Shuo Chen, Qisen Yang, Andrew Zhao, Chaofei Wang, Shiji Song, and Gao Huang.
\newblock Boosting llm agents with recursive contemplation for effective deception handling.
\newblock In {\em The 62nd Annual Meeting of the Association for Computational Linguistics}, 2024.

\bibitem{wangGEMINIFastFailure2023a}
Zhuang Wang, Zhen Jia, Shuai Zheng, Zhen Zhang, Xinwei Fu, T.~S.~Eugene Ng, and Yida Wang.
\newblock {{GEMINI}}: {{Fast Failure Recovery}} in {{Distributed Training}} with {{In-Memory Checkpoints}}.
\newblock In {\em Proceedings of the 29th {{Symposium}} on {{Operating Systems Principles}}}, {{SOSP}} '23, pages 364--381, New York, NY, USA, October 2023. Association for Computing Machinery.

\bibitem{weiNoProvisionedConcurrency2023b}
Xingda Wei, Fangming Lu, Tianxia Wang, Jinyu Gu, Yuhan Yang, Rong Chen, and Haibo Chen.
\newblock No {{Provisioned Concurrency}}: {{Fast}} \{\vphantom\}{{RDMA-codesigned}}\vphantom\{\} {{Remote Fork}} for {{Serverless Computing}}.
\newblock In {\em 17th {{USENIX Symposium}} on {{Operating Systems Design}} and {{Implementation}} ({{OSDI}} 23)}, pages 497--517, 2023.

\bibitem{wolf-etal-2020-transformers}
Thomas Wolf, Lysandre Debut, Victor Sanh, Julien Chaumond, Clement Delangue, Anthony Moi, Pierric Cistac, Tim Rault, R{\'e}mi Louf, Morgan Funtowicz, Joe Davison, Sam Shleifer, Patrick {von Platen}, Clara Ma, Yacine Jernite, Julien Plu, Canwen Xu, Teven~Le Scao, Sylvain Gugger, Mariama Drame, Quentin Lhoest, and Alexander~M. Rush.
\newblock Transformers: {{State-of-the-art}} natural language processing.
\newblock In {\em Proceedings of the 2020 Conference on Empirical Methods in Natural Language Processing: {{System}} Demonstrations}, pages 38--45, Online, October 2020. Association for Computational Linguistics.

\bibitem{wuLoongServeEfficientlyServing2024a}
Bingyang Wu, Shengyu Liu, Yinmin Zhong, Peng Sun, Xuanzhe Liu, and Xin Jin.
\newblock {{LoongServe}}: {{Efficiently Serving Long-Context Large Language Models}} with {{Elastic Sequence Parallelism}}.
\newblock In {\em Proceedings of the {{ACM SIGOPS}} 30th {{Symposium}} on {{Operating Systems Principles}}}, {{SOSP}} '24, pages 640--654, New York, NY, USA, November 2024. Association for Computing Machinery.

\bibitem{wuDLoRADynamicallyOrchestrating2024}
Bingyang Wu, Ruidong Zhu, Zili Zhang, Peng Sun, Xuanzhe Liu, and Xin Jin.
\newblock \{\vphantom\}{{dLoRA}}\vphantom\{\}: {{Dynamically Orchestrating Requests}} and {{Adapters}} for \{\vphantom\}{{LoRA}}\vphantom\{\} \{\vphantom\}{{LLM}}\vphantom\{\} {{Serving}}.
\newblock In {\em 18th {{USENIX Symposium}} on {{Operating Systems Design}} and {{Implementation}} ({{OSDI}} 24)}, pages 911--927, 2024.

\bibitem{yangOndemandParallelCheckpoint2024}
Yanning Yang, Dong Du, Haitao Song, and Yubin Xia.
\newblock On-demand and {{Parallel Checkpoint}}/{{Restore}} for {{GPU Applications}}.
\newblock In {\em Proceedings of the 2024 {{ACM Symposium}} on {{Cloud Computing}}}, {{SoCC}} '24, pages 415--433, New York, NY, USA, November 2024. Association for Computing Machinery.

\bibitem{yaoTreeThoughtsDeliberate2023}
Shunyu Yao, Dian Yu, Jeffrey Zhao, Izhak Shafran, Tom Griffiths, Yuan Cao, and Karthik Narasimhan.
\newblock Tree of {{Thoughts}}: {{Deliberate Problem Solving}} with {{Large Language Models}}.
\newblock {\em Advances in Neural Information Processing Systems}, 36:11809--11822, December 2023.

\bibitem{yuOrcaDistributedServing2022}
Gyeong-In Yu, Joo~Seong Jeong, Geon-Woo Kim, Soojeong Kim, and Byung-Gon Chun.
\newblock Orca: {{A Distributed Serving System}} for \{\vphantom\}{{Transformer-Based}}\vphantom\{\} {{Generative Models}}.
\newblock In {\em 16th {{USENIX Symposium}} on {{Operating Systems Design}} and {{Implementation}} ({{OSDI}} 22)}, pages 521--538, 2022.

\bibitem{zhangOPTOpenPretrained2022}
Susan Zhang, Stephen Roller, Naman Goyal, Mikel Artetxe, Moya Chen, Shuohui Chen, Christopher Dewan, Mona Diab, Xian Li, Xi~Victoria Lin, Todor Mihaylov, Myle Ott, Sam Shleifer, Kurt Shuster, Daniel Simig, Punit~Singh Koura, Anjali Sridhar, Tianlu Wang, and Luke Zettlemoyer.
\newblock {{OPT}}: {{Open Pre-trained Transformer Language Models}}, June 2022.

\bibitem{zhongDistServeDisaggregatingPrefill2024}
Yinmin Zhong, Shengyu Liu, Junda Chen, Jianbo Hu, Yibo Zhu, Xuanzhe Liu, Xin Jin, and Hao Zhang.
\newblock \{\vphantom\}{{DistServe}}\vphantom\{\}: {{Disaggregating Prefill}} and {{Decoding}} for {{Goodput-optimized Large Language Model Serving}}.
\newblock In {\em 18th {{USENIX Symposium}} on {{Operating Systems Design}} and {{Implementation}} ({{OSDI}} 24)}, pages 193--210, 2024.

\end{thebibliography}

\end{document}